\documentclass[aps,floatfix,eqsecnum,showpacs,preprint,tightenlines]{revtex4-1}
\usepackage{epsfig}
\usepackage{bm}
\usepackage{amsmath}

\def\e{\kern+.5ex\lower.42ex\hbox{$\scriptstyle \iota$}\kern-1.10ex e}
\def\registered{{\ooalign{\hfil\raise .00ex\hbox{\scriptsize R}\hfil\crcr\mathhexbox20D}}}

\newcommand{\BA}[1]{\langle #1 \mid}

\newcommand{\KT}[1]{\mid #1 \rangle}

\begin{document}

%\preprint{}

\title{Momentum space treatment of inclusive neutrino scattering off the deuteron and trinucleons}


%\author{}
%\thanks{}
%\affiliation{}

\author{J. Golak}
\affiliation{M. Smoluchowski Institute of Physics, Jagiellonian University, PL-30348 Krak\'ow, Poland}
\author{R. Skibi{\'n}ski}
\affiliation{M. Smoluchowski Institute of Physics, Jagiellonian University, PL-30348 Krak\'ow, Poland}
\author{K. Topolnicki}
\affiliation{M. Smoluchowski Institute of Physics, Jagiellonian University, PL-30348 Krak\'ow, Poland}
\author{H. Wita{\l}a}
\affiliation{M. Smoluchowski Institute of Physics, Jagiellonian University, PL-30348 Krak\'ow, Poland}
\author{A. Grassi}
\affiliation{M. Smoluchowski Institute of Physics, Jagiellonian University, PL-30348 Krak\'ow, Poland}
\author{H. Kamada}
\affiliation{Department of Physics, Faculty of Engineering,
Kyushu Institute of Technology, Kitakyushu 804-8550, Japan}
\author{L.E. Marcucci}
\affiliation{Department of Physics, University of Pisa, IT-56127 Pisa, Italy
              and INFN-Pisa, IT-56127 Pisa, Italy}


\date{\today}

\begin{abstract}
The 
$\bar{\nu}_e + {^2{\rm H}} \rightarrow e^+ + n + n$, 
$\nu_e + {^2{\rm H}} \rightarrow e^- + p + p$, 
$\bar{\nu}_l + {^2{\rm H}} \rightarrow \bar{\nu}_l + {^2{\rm H}}$,
$\nu_l + {^2{\rm H}} \rightarrow \nu_l + {^2{\rm H}}$,
$\bar{\nu}_l + {^2{\rm H}} \rightarrow \bar{\nu}_l + p + n$, 
$\nu_l + {^2{\rm H}} \rightarrow \nu_l + p + n$,
$\bar{\nu}_e + {^3{\rm He}} \rightarrow e^+ + {^3{\rm H}}$, 
$\bar{\nu}_l + {^3{\rm He}} \rightarrow \bar{\nu}_l + {^3{\rm He}}$, 
$\nu_l + {^3{\rm He}} \rightarrow \nu_l + {^3{\rm He}}$,
$\bar{\nu}_l + {^3{\rm H}} \rightarrow \bar{\nu}_l + {^3{\rm H}}$, 
$\nu_l + {^3{\rm H}} \rightarrow \nu_l + {^3{\rm H}}$,
$\bar{\nu}_e + {^3{\rm He}} \rightarrow e^+ + n + d$, 
$\bar{\nu}_e + {^3{\rm He}} \rightarrow e^+ + n + n + p$, 
$\bar{\nu}_l + {^3{\rm He}} \rightarrow \bar{\nu}_l + p + d$, 
$\bar{\nu}_l + {^3{\rm He}} \rightarrow \bar{\nu}_l + p + p +n$, 
$\nu_l + {^3{\rm H}} \rightarrow \nu_l + n + d$
and
$\nu_l + {^3{\rm H}} \rightarrow \nu_l + n + n + p$
reactions ($l= e, \mu, \tau$) are studied consistently in momentum space for (anti)neutrino energies
up to 300 MeV.
For most of these processes we provide predictions for the total cross sections
and in the case of the (anti)neutrino-$^3$He 
and (anti)neutrino-$^3$H inelastic scattering we compute examples of essential 
response functions, using the AV18 nucleon-nucleon potential and 
a single-nucleon weak current operator.
For the reactions with the deuteron 
we study relativistic effects in the final state kinematics 
and compare two-nucleon scattering states obtained 
in momentum and coordinate spaces.
Our results from momentum space are compared
with the theoretical predictions by
G.~Shen {\it et al.}, Phys. Rev. C{\bf 86}, 035503 (2012).
The observed disagreement can be attributed to the differences 
in kinematics and in the weak current operator.
\end{abstract}

\pacs{23.40.-s, 21.45.-v, 27.10.+h}
%\keywords{}

\maketitle


\section{Introduction}
\label{section1}

Neutrino scattering on nuclei has been studied for several decades.
For information on the status of earlier theoretical treatments of neutrino-nucleus reactions,
relevant to the detection of astrophysical neutrinos, we refer the reader to Ref.~\cite{Kubodera94}.
At the beginning of the century theoretical work was motivated by the establishment of the 
Sudbury Neutrino Observatory and resulted in important predictions by
Nakamura {\it et al.} \cite{PRC63.034617,NPA707.561}. They calculated 
cross sections for both charged-current (CC) and neutral-current (NC) driven reactions,
for incoming neutrino energies up to 150 MeV. 
The results of Ref.~\cite{PRC63.034617} and the bulk of predictions given
in Ref.~\cite{NPA707.561} were
obtained within the so-called "standard nuclear physics approach" \cite{Carlson98},
employing the AV18 nucleon-nucleon (NN) force \cite{AV18} and supplementing the single-nucleon current
with two-nucleon (2N) current contributions related to this potential. Some calculations 
in Ref.~\cite{NPA707.561} were done with the CD-Bonn NN potential~\cite{CDBonn} 
or using input from chiral effective field theory ($\chi$EFT) in order to estimate 
theoretical uncertainties of the results, which were later used to analyze 
experimental data from the Sudbury Neutrino Observatory~\cite{Ahmad02}.

More recent calculations by Shen {\it et al.}~\cite{PRC86.035503} were also 
done within the traditional approach, using the AV18 potential 
and corresponding nuclear weak current operators with a one-body part and two-body 
contributions, adjusted to the NN force.
The authors of Ref.~\cite{PRC86.035503} studied inclusive neutrino scattering on the
deuteron up to neutrino energies of 1 GeV with configuration space methods. 
Although they introduced some
changes in the 2N current operator used by  Nakamura {\it et al.}, 
these modifications proved to be of minor importance 
and the results obtained by Shen {\it et al.} confirmed those 
of Nakamura {\it et al.} in the energy range up to 150 MeV.
Conclusions presented in Ref.~\cite{PRC86.035503} provided important
information on the role of 2N currents and final state interaction
effects for the whole considered neutrino energy range, even though 
pion production channels were neglected.

Despite the successes achieved within the traditional approach, 
new calculations emerging from $\chi$EFT 
offered competitive results. Already in 2001  
Butler {\it et al.}  \cite{Butler2001} studied 
the neutrino-deuteron break-up reactions at next-to-next-to-leading order (N2LO)
in pionless $\chi$EFT,
in the energy range up to 20 MeV. Their work agreed very well
with the previous potential model calculations from 
Refs.~\cite{PRC63.034617,NPA707.561}.

Attempts to build a complete theoretical framework comprising consistent 
``chiral'' 2N and many-nucleon forces as well as electroweak current 
operators at a sufficiently high order of the chiral expansion
have a long history. 
A construction of the chiral NN potential 
was pioneered by Weinberg~\cite{Weinberg1,Weinberg2} 
almost thirty years ago and developed by several groups. 
In particular Epelbaum {\it et al.} have prepared three generations
of the chiral potentials. They started with the version 
of the NN potential, where the non-local 
regularization in momentum space was 
implemented~\cite{Epelbaum_start,SFR,nucleon-nucleon-N3LO}.
They derived also the widely used chiral three-nucleon (3N) potential at N2LO~\cite{Epelbaum_3N},
summarizing the work on chiral forces and their applications 
to processes involving few nucleons up to 2005 in Ref.~\cite{Epelbaum2006}.
Further important contributions from this group dealt with 
the 3N force at next-to-next-to-next-to-leading order (N3LO)
\cite{Bernard1-3NF,Bernard2-3NF},
the four-nucleon force~\cite{EE4Nforce},
and a formulation of the $\Delta$-full chiral perturbation theory
\cite{Delta-Krebs,Delta-Krebs2}.

The next generation of the chiral NN potential by Epelbaum {\it et al.}
used a coordinate space regularization. This improved version from Refs.~\cite{imp1,imp2}
led to a significant reduction of finite-cutoff artefacts,
did not require any additional spectral function regularization 
and directly employed low-energy constants determined from pion-nucleon scattering.
These forces were used to study nucleon-deuteron scattering \cite{Binder}
and various electroweak processes in 2N and 3N systems \cite{Skibinski2016}.

The newest version of the Bochum-Bonn 
chiral NN potential, prepared up to fifth order in the chiral expansion (N4LO),
was introduced very recently in Ref.~\cite{Reinert}. 
Important changes include a removal of the redundant contact terms and
regularization in momentum space, resulting in
an excellent description of the proton-proton and
neutron-proton scattering data from the self-consistent Granada-2013 database~\cite{Granada2013}.

The Bochum-Bonn group has also been working on the chiral electromagnetic
\cite{Kolling1,Kolling2} and weak (axial) \cite{Krebs2017}  current operators.
First results with the 2N electromagnetic currents from Ref.~\cite{Kolling1}
were published in Refs.~\cite{Rozpedzik,Skibinski_APP2}, but full-fledged calculations 
with the consistent Bochum-Bonn potentials and current operators will become possible, 
when the ongoing work on the regularization of the current operators is completed. 

Concurrent with these studies have been the 
efforts by the Moscow(Idaho)-Salamanca group, which resulted in another 
family of non-local chiral NN potentials \cite{Entem2003,Machleidt2011}.
The most recent version of this potential, generated also up to fifth order in the chiral expansion
was published in Ref.~\cite{Machleidt2017}. 
At N4LO it reproduces the world NN data with the $\chi^2$/datum of 1.08 for 
proton-proton and neutron-proton data up to 190 MeV.

Many modern calculations of various electromagnetic processes employ
the chiral potentials from Refs.~\cite{Entem2003,Machleidt2011} and require 
chiral current operators. The latter were developed gradually, starting with a pioneering work 
by Park {\it et al.}~\cite{Park93}. The predictions of Refs.~\cite{Park93,Park96} were later re-derived 
or supplemented by many authors \cite{Pastore08,Pastore09,Pastore11,Baroni16},
using various formulations of $\chi$EFT. The unknown parameters
of the effective theory were either related to the NN scattering
or fixed by reproducing selected observables in the 2N and 3N systems, like 
the magnetic moments \cite{Piarulli13} and the tritium Gamow-Teller matrix 
element~\cite{Gazit09,Baroni16a}.
The derived current operators were used with the wave functions 
obtained with the traditional potentials and later, more consistently, 
with the potentials derived by the Moscow(Idaho)-Salamanca group.
Among the many studied processes were those of direct astrophysical 
interest~\cite{Park03,Marcucci13}, muon capture reactions ~\cite{Gazit08,Marcucci11,Marcucci12,Marcucci14}
and, last but not least, neutrino induced processes~\cite{Baroni17}.

Predictions in Ref.~\cite{Baroni17} for inclusive neutrino scattering off the deuteron
are fully based on a $\chi$EFT input.
The results concerning the cross sections are only slightly larger 
than the corresponding ones obtained in conventional formulations based on meson-
exchange picture \cite{NPA707.561,PRC86.035503}
and are insensitive to the value of the regulator parameter. This
might indicate that the theoretical results have a very small uncertainty
in the low-energy neutrino regime.

To give the reader a proper picture of the efforts aiming at the
exact treatment of the neutrino induced reactions, we mention here 
some calculations with heavier than $A=2$ nuclei. 
Gazit {\it et al.} performed a number of calculations for neutrino induced break-up 
reactions with the $^3$H, $^3$He and $^4$He nuclei~\cite{Gazit04,Gazit07,Gazit07a},
in which final state interactions were included via the Lorentz integral transform method ~\cite{LITM}.
The resulting bound-state and bound-state-like equations were solved using the effective 
interaction hyperspherical harmonics (EIHH) approach~\cite{EIHH1,EIHH2}, employing conventional 
2N and 3N forces.
While in Ref.~\cite{Gazit04} the impulse approximation was used, in Refs.~\cite{Gazit07,Gazit07a}
the nuclear current operator contained also 2N contributions derived 
from $\chi$EFT.
Finally, we mention that weak inclusive responses of heavier light nuclei, 
including $^{12}$C, were investigated with the Green's function Monte Carlo 
method~\cite{Lovato2014,Lovato2015}. The results of these calculations contributed to 
the determination of the nucleon isovector axial form factor ~\cite{Meyer2016}.

The momentum-space approach offers an independent 
possibility to perform calculations not only for the deuteron but also for the 
trinucleons' reactions with neutrinos. In the present work we calculate cross sections
for several such reactions and build a solid base upon which we can improve our dynamics in the future,
adding many-nucleon forces and weak current operators. 
The present study, contrary to the very advanced investigations 
by Baroni {\it et al.} \cite{Baroni17}, is carried out with rather simple dynamical input.
Namely we work with the traditional AV18 NN potential and restrict ourselves 
to the single nucleon current. Thus we definitely cannot reach yet the high level of 
accuracy achieved by the predictions of Refs.~\cite{NPA707.561,PRC86.035503,Baroni17}, dealing solely with
the neutrino induced break-up of $^2$H. We agree with the statement in Ref.~\cite{Baroni17} that 
the accuracy of these predictions is very important in the analysis of the SNO experiments 
and more generally for our understanding of (anti)neutrino-nucleus scattering. 
Thus we decided to confront our momentum-space framework predictions
with the above-mentioned results. 
We also agree with Ref.~\cite{Baroni17}
that all these ingredients should be derived consistently from $\chi$EFT. 
There are, however, still some open issues in the construction of the 2N
electroweak current operator and the results for the axial current 
obtained by Krebs {\it et al.}~\cite{Krebs2017} 
are not equivalent to those reported in Refs.~\cite{Baroni16,Baroni16a}.
Even if these differences prove to be of no practical importance,
some fundamental questions about the consistence between chiral 
potentials and current operators still should be answered.  
Our framework is anyway ready for the improved input generated by $\chi$EFT.

The results obtained within $\chi$EFT are usually provided in momentum space
and can be readily incorporated in momentum-space calculations.
We refer the reader especially to the so-called "three-dimensional" calculations, 
which avoid totally partial wave representation of nuclear states 
and operators~\cite{Topolnicki13,PRC90.024001}.
In this approach $\chi$EFT potentials and current operators 
would be used indeed directly, avoiding also convergence problems
bound with partial wave decomposition. Our present results might thus provide
a benchmark for such planned calculations.

Last but not least, momentum space framework allows one to 
systematically account for relativistic effects not only 
in the kinematics but also in the reaction dynamics.
Some of such problems might be difficult to tackle in coordinate space
but are easier to solve in momentum space.
For example, the argument 
of the nucleon form factors in the single-nucleon current, which 
should be actually the four-momentum transfer {\em to the nucleon} 
squared, is usually replaced by the four-momentum transfer 
{\em to the whole nuclear system} squared. 
In momentum space one can directly use the proper values
of the form factor arguments.
  
Relativity plays definitely an important role for higher 
neutrino energies. Even at relatively low energies these effects have to 
be studied thoroughly, since approximate treatment of relativity 
adds to the total theoretical uncertainty of predictions.
Ultimately, theoretical calculations should take into account
complementary roles of kinematic and dynamical contributions 
to the Poincar\'e invariant formulation of reaction theory.
Important examples of such investigations are given in 
Refs.~\cite{Kamada02,Witala05,Golak06,Liu08,Witala11,Polyzou11,Polyzou14}.
A particular result of these studies, showing that relativistic effects in
kinematics and dynamics might in fact partly cancel, 
prompted us to retain the nonrelativistic form of the phase space factor,
consistent with our nonrelativistic dynamics, in particular with the form 
of the current operator. Consequences of this choice will be discussed below.

The paper is organized in the following way.
In Sec.~\ref{section2} we introduce elements 
of our formalism and compare it with the 
calculations presented in Refs.~\cite{PRC63.034617,PRC86.035503}. 
In particular we discuss the differences 
in the treatment of kinematics and the 
current operator.
In the following two sections we show selected results for 
various processes induced by neutrinos.
Finally, Sec.~\ref{section7} contains some concluding remarks and outlook.

\section{Elements of the formalism}
\label{section2}

Recently we have developed a framework to study several muon capture 
processes on the $^2{\rm H}$, $^3{\rm H}$ and $^3{\rm H}$ 
nuclei~\cite{PRC90.024001,PRC94.034002}. 
For the
$\bar{\nu}_l + {^2{\rm H}} \rightarrow l^+ + n + n$ reaction,
the transition from the initial to final state is also governed by the
Fermi form of the interaction Lagrangian~\cite{walecka},
again leading to a contraction of the leptonic (${\cal L}_\lambda$) and nuclear
(${\cal N}^\lambda$) parts in the $S$~matrix element, $S_{fi}$:
\begin{eqnarray}
S_{fi}= i ( 2 \pi )^4 \, \delta^4 \left( P^\prime - P \right)\, 
\frac {G_F \, \cos\theta_C}{\sqrt{2}} \, {\cal L}_\lambda \, {\cal N}^\lambda \, ,
\label{sfi}
\end{eqnarray}
where the value of the Fermi constant, $G_F = 1.1803 \times 10^{-5} \, {\rm GeV}^{-2} $, 
and $\cos\theta_C = 0.97425$ have been deliberately taken to be the same as in 
Ref.~\cite{PRC86.035503}. The total initial (final) four-momentum is denoted 
as $P$ ($P^\prime$).

The leptonic matrix element
\begin{eqnarray}
{\cal L}_\lambda = \frac 1{ \left( 2 \pi \, \right)^3 } \, 
\bar{v} ( {\bf k} , m_{\bar{\nu}} ) \gamma_\lambda ( 1- \gamma_5 ) 
v ( {\bf k}^\prime , m_{l^+} )
\, \equiv \,
\frac 1{ \left( 2 \pi \, \right)^3 } \, L_\lambda
\label{llambda}
\end{eqnarray}
is given in terms of the Dirac spinors $v$ and the gamma matrices \cite{bjodrell}, and depends
on the initial antineutrino three-momentum ${\bf k}$ and
spin projection $m_{\bar{\nu}}$ as well as
on the final antilepton three-momentum ${\bf k}^\prime$ and 
spin projection $m_{l^+}$. The same formula holds for the three lepton flavors $l=e$, $l=\mu$ and $l=\tau$.

The nuclear part
\begin{eqnarray}
{\cal N}^\lambda = \frac 1{ \left( 2 \pi \, \right)^3 } \, 
\BA{\Psi_f \, {\bf P}_f \, m_{f} \, } \, 
j_{CC}^\lambda
\, \KT{\Psi_i \, {\bf P}_i \, m_{i} \, } 
\, \equiv \,
\frac 1{ \left( 2 \pi \, \right)^3 } \, N^\lambda_{CC}
\label{nlambda}
\end{eqnarray}
is a matrix element of the nuclear weak charged current (CC) operator 
$j_{CC}^\lambda$ between the initial and final nuclear states.
The total initial (final) nuclear three-momentum is denoted as ${\bf P}_i$ (${\bf P}_f$),
$m_i$ is the initial nucleus spin projection and $m_f$ is the set of spin projections
in the final state. In this paper we restrict ourselves to the single nucleon current 
operator with relativistic corrections. This current operator was described in detail 
in Ref.~\cite{PRC90.024001}. It is very close to the one used in Ref.~\cite{Marcucci11}
and employs form factors, whose explicit expressions and parametrization can be found
in Ref.~\cite{PRC86.035503}.

On top of the single nucleon operators,
also many-nucleon contributions appear in $j_{CC}^\lambda$.
Their role has been studied 
for example in Ref.~\cite{Marcucci11}.
For the neutrino induced reactions of interest, the effects of 2N 
contributions in the weak current operator were estimated in Ref.~\cite{PRC86.035503}
to be smaller than $10 \%$ over the wide energy range from the threshold to GeV energies.
Thus we decided to base our first predictions on the single nucleon current only
and represent all dynamical ingredients in momentum space.

The only change 
in the charged single nucleon current operator
for the ${\nu}_l + {^2{\rm H}} \rightarrow l^- + p + p$
process compared to the 
$\bar{\nu}_l + {^2{\rm H}} \rightarrow l^+ + n + n$ reaction
is the replacement of the overall isospin lowering operator
by the isospin raising operator:
\begin{eqnarray}
{\tau}_{-} \equiv(\tau_x -{\rm i} \tau_y)/2 \longrightarrow {\tau}_{+} \equiv(\tau_x + {\rm i} \tau_y)/2 \, .
\label{isospin}
\end{eqnarray}
However, since the matrix elements of the single nucleon operator 
in the 2N isospin space, spanned by the $ \mid \big( \frac12 \frac12 \big) t m_t  \, \big\rangle $ states,
$\big\langle  \big( \frac12 \frac12 \big) 1 -1 \mid \tau_{-} (1) \mid \big( \frac12 \frac12 \big) 0  0 \, \big\rangle $
and
$\big\langle  \big( \frac12 \frac12 \big) 1  1 \mid \tau_{+} (1) \mid \big( \frac12 \frac12 \big) 0  0 \, \big\rangle $
have just an opposite sign, we can use for this reaction {\em exactly} the same single nucleon current operator.
Its matrix elements, $N^\lambda_{CC}$, are contracted with the altered  
leptonic matrix elements
\begin{eqnarray}
L_\lambda = 
\bar{u} ( {\bf k}^\prime , m_{l^-} ) \gamma_\lambda ( 1- \gamma_5 ) 
u ( {\bf k} , m_{\nu} ) \, ,
\label{llambdapp}
\end{eqnarray}
expressed through the Dirac spinors $u$, which depend
on the initial neutrino three-momentum ${\bf k}$ and
spin projection $m_{\nu}$ as well as
on the final lepton three-momentum ${\bf k}^\prime$ and 
spin projection $m_{l^-}$. In the following the energy of the initial 
(anti)neutrino will be denoted by ${ E}$ and for the massless (anti)neutrino 
$ E = \mid {\bf k} \mid $.

\subsection{Kinematics}
\label{section2.1}

Since we compare our purely nonrelativistic predictions 
with the ones published in
Ref.~\cite{PRC86.035503}, where the relativistic kinematics was employed,
we give here formulas for our kinematics and
cross sections for all studied reactions. We believe that they will
be useful in the future benchmark calculations and serve to  
disentangle relativistic kinematical effects from dynamical ones.
The difference in the treatment of kinematics is the main reason, why 
our predictions diverge from the results published in Ref.~\cite{PRC86.035503},
especially for higher energies.

The kinematics of the 
$\bar{\nu}_l + {^2{\rm H}} \rightarrow l^+ + n + n$ and
${\nu}_l + {^2{\rm H}} \rightarrow l^- + p + p$ 
processes is essentially identical (the only difference being the 
mass of two identical nucleons in the final state) and can be treated 
both relativistically and nonrelativistically. 
The relativistic formulas for different kinematical quantities
are given in Refs.~\cite{PRC63.034617,PRC86.035503}, so we can focus
on the differences between the exact relativistic 
and our approximate nonrelativistic treatment of kinematics.

The starting point is the energy and momentum conservation,
where we neglect the very small (anti)neutrino mass 
and assume that the initial deuteron is at rest. 
In the relativistic formalism it reads:
\begin{eqnarray} 
%\mid {\bf k} \mid 
{ E} + M_d &=&  
\sqrt{ M_l^2 + {{\bf k}^{\prime}}^{\, 2} \, }  
+ 
\sqrt{ M_p^2 + {\bf p}_1^{\, 2} \, }   
+
\sqrt{ M_p^2 + {\bf p}_2^{\, 2} \, }  \, , \nonumber  \\
{\bf k} &=& {\bf k}^\prime + {\bf p}_1 + {\bf p}_2 \, .
\label{relnn}
\end{eqnarray}  
Here ${\bf p}_1$ and ${\bf p}_2$ stand for the individual momenta of the two outgoing nucleons.
The deuteron, nucleon and (anti)lepton masses are denoted as
$M_d$, $M_p$ and $M_l$, respectively.
In the nonrelativistic version of Eqs.~(\ref{relnn}) we replace the first equation by
\begin{eqnarray} 
%\mid {\bf k} \mid 
{ E} + M_d =
\sqrt{ M_l^2 + {{\bf k}^{\prime}}^{\, 2} \, }  
+ 
2 M_p + 
\frac { {\bf p}_1^{\, 2} \, }  { 2 M_p}  
+
\frac { {\bf p}_2^{\, 2} \, }  { 2 M_p} \, ,
\label{nrlnn}
\end{eqnarray}  
using nonrelativistic formulas in the nuclear sector.
We make sure to what extent the nonrelativistic approximation is justified 
by comparing values of various quantities calculated nonrelativistically 
and using relativistic equations. This is important, since our dynamics 
is entirely nonrelativistic.

We begin with the (anti)neutrino threshold energy, ${ E}_{thr}$.
Relativistically, the condition for the total energy of the system in the zero-momentum 
frame
\begin{eqnarray} 
\left( 
%\mid {\bf k} \mid 
{ E} + M_d \, \right)^2 - {\bf k}^{2} \ge \left( M_l + 2 M_p \,  \right)^2
\label{relkthr}
\end{eqnarray}
leads to
\begin{eqnarray} 
%\mid {\bf k} \mid  
{ E} \ge { E}_{thr}^{rel} \equiv \frac{  \left( M_l + 2 M_p \,  \right)^2 - M_d^2 }{2 M_d} \, .
\label{relkthr2}
\end{eqnarray}

In the fully nonrelativistic approach, one treats even the final (anti)lepton 
nonrelativistically and obtains 
\begin{eqnarray} 
%\mid {\bf k} \mid 
{ E} + M_d =
M_l + 2 M_p +
\frac { {{\bf k}^{\prime}}^{\, 2}  }{ 2 M_l}  
+
\frac { {\bf p}_1^{\, 2} \, }  { 2 M_p}  
+
\frac { {\bf p}_2^{\, 2} \, }  { 2 M_p} \, \equiv \, 
M_l + 2 M_p + T_{lab} \, ,
\label{nrlnn2}
\end{eqnarray}  
where $T_{lab}$ is the total kinetic energy in the laboratory frame. 
Then the threshold energy is obtained from the condition that the kinetic energy 
calculated in the center of mass frame, $T_{cm}$, is non-negative:
\begin{eqnarray}
T_{cm} = T_{lab}  - \frac { {{\bf k}}^{\, 2}  }{ 2 \left( M_l + 2 M_p \, \right) } \ge 0 
\label{nrlnn3}
\end{eqnarray}
and reads
\begin{eqnarray} 
{ E}_{thr}^{nrl}=  M_l + 2 M_p -\sqrt{  \left( M_l + 2 M_p \, \right) \, \left( 2 M_d - M_l - 2 M_p \, \right) \, } \, .
\label{nrlnn4}
\end{eqnarray}
Actually this result does not follow from Eq.~(\ref{nrlnn}), where we apply the nonrelativistic 
formulas only to the outgoing nucleons. In order to be consistent, we rewrite Eq.~(\ref{nrlnn}) as
\begin{eqnarray}
%\mid {\bf k} \mid 
{ E} + M_d =
\sqrt{ M_l^2 + {{\bf k}^{\prime}}^{\, 2} \, }  
+ 2 M_p +
\frac { {\bf p}_{12}^{\, 2} \, }  { 4 M_p}  
+
\frac { {\bf p}^{\, 2} \, }  { M_p} \, ,
\label{nrlnn5}
\end{eqnarray}  
where $  {\bf p}_{12} \equiv  {\bf p}_{1} +  {\bf p}_{2} =  {\bf k} - {\bf k}^{\prime} $ and 
$ {\bf p} = \frac12 \left(  {\bf p}_{1} -  {\bf p}_{2} \, \right) $.
Next we numerically seek 
the smallest possible value of $ { E} $ (represented by ${\cal E}_{thr}^{nrl}$), for which
a physical solution of %for $\mid {\bf k}^{\prime} \mid $
\begin{eqnarray}
%\mid {\bf k} \mid 
{ E} + M_d =
\sqrt{ M_l^2 + {{\bf k}^{\prime}}^{\, 2} \, }  
+ 2 M_p +
\frac { \left(  {\bf k} - {\bf k}^{\prime}  \, \right)^2   \, }  { 4 M_p}  
\label{nrlnn6}
\end{eqnarray}  
exists.

Obviously the same logic applies to the $ \nu_l + d \rightarrow \nu_l  + p + n $
and
$ \bar{\nu}_l + d \rightarrow \bar{\nu}_l  + p + n $ 
reactions. In this case the relativistic result is still exact,
but in the nonrelativistic calculations we additionally 
neglect the small difference between the proton mass $M_p$
and neutron mass $M_n$
and use the average ``nucleon mass'',
$M \equiv \frac12 \left( M_p + M_n \, \right) $.
For the massless particle in the final state, the fully nonrelativistic 
calculation is not possible.
From Table~\ref{tabkthr}, where we display all the numerical results 
for the threshold energies, it is clear that difference between the
relativistic and nonrelativistic results are insignificant.

\begin{table}%[H] add [H] placement to break table across pages
\caption{Threshold energies in MeV for various (anti)neutrino induced reactions 
on the deuteron calculated relativistically (${ E}_{thr}^{rel}$)
and using two nonrelativistic prescriptions (${ E}_{thr}^{nrl}$ and ${\cal E}_{thr}^{nrl}$).
We assumed 
$ M_d$ = 1875.613 MeV,
$ M_p$ = 938.272 MeV,
$ M_n$ = 939.565 MeV
%$M_\mu$ = 105.658 MeV
and 
$M_e$ = 0.510999 MeV.
Results for the $ \nu_l + d \rightarrow \nu_l  + p + n $ 
and $ \bar{\nu}_l + d \rightarrow \bar{\nu}_l  + p + n $
processes are identical.
\label{tabkthr}}
\begin{tabular}{c|c|c|c}
\hline\hline
reaction & ${ E}_{thr}^{rel}$  &  ${ E}_{thr}^{nrl}$  & ${\cal E}_{thr}^{nrl}$  \\
\hline
$ \bar{\nu}_e + d \rightarrow e^+ + n + n $      &  4.03323  &  4.03323 & 4.03323 \\
$ \nu_e + d \rightarrow e^- + p + p $            &  1.44279  &  1.44279 & 1.44279 \\

$ \nu_l + d \rightarrow \nu_l  + p + n $         &  2.22589  &     --   & 2.22589 \\
\hline\hline
\end{tabular}
\end{table}

Next, we determine the maximal energy of the emerging lepton under a given
scattering angle $\theta$, where $\cos\theta = \hat{\bf k} \cdot {\hat{\bf k}}^{\, \prime}$. 
Note that there is no restriction on the scattering angle $\theta$. 
Relativistically, we set the condition for the total energy of two nucleons in their zero-momentum 
frame 
\begin{eqnarray} 
\left(
%\mid {\bf k} \mid 
{ E} + M_d - \sqrt{ M_l^2 + {{\bf k}^{\prime}}^{\, 2} \, }  \, \right)^2 - 
\left( {\bf k} - {\bf k}^{\prime} \, \right)^2 \ge 4 M_p^2 \, .
\label{relnn3}
\end{eqnarray}  
Simple algebra leads just to a quadratic inequality for $k^\prime \equiv \mid {\bf k}^{\prime} \mid$:
\begin{eqnarray} 
\left( 4 ( { E} + M_d)^2 - 4 { E}^2 \cos\theta \right) \, {k^\prime}^{\, 2} \nonumber \\
+
\left( -4 { E} (2 { E} M_d + M_d^2 + M_l^2 - 4 M_p^2) \cos\theta \, \right) \, {k^\prime} \nonumber \\
+
4 ({ E} + M_d)^2 M_l^2 - \left(2 { E} M_d + M_d^2 + M_l^2 - 4 M_p^2 \right)^2
\le 0 
\label{relnn4}
\end{eqnarray}  
and the bigger of the two roots of the corresponding quadratic equation is 
the maximal value of the magnitude of the outgoing lepton, $ \left( k^\prime \right)_{max}^{rel} $.

\begin{figure}
\includegraphics[width=7cm]{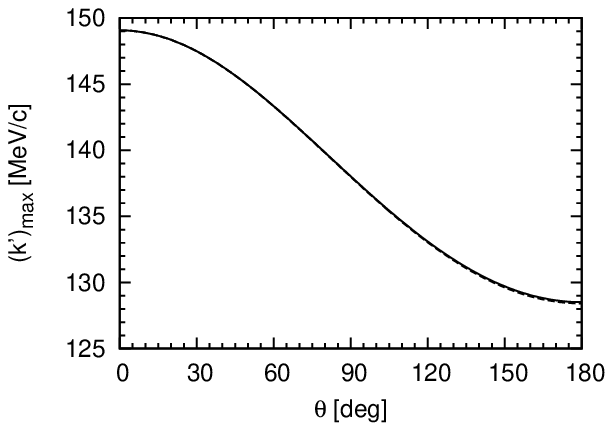}
\includegraphics[width=7cm]{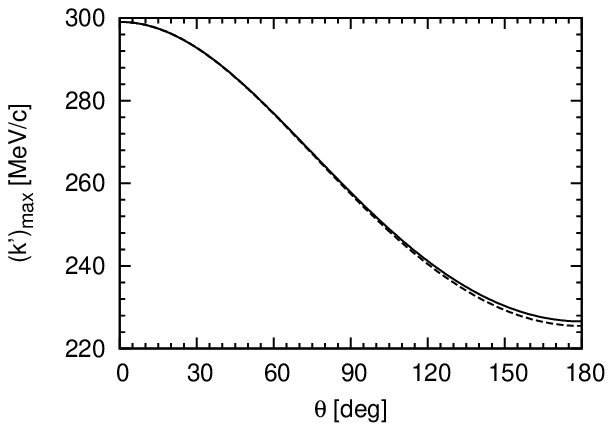}
\caption{Maximal value of the outgoing electron momentum $k^\prime$ 
in the 
${\nu}_e + {^2{\rm H}} \rightarrow e^- + p + p$ reaction
as a function of the lepton scattering angle $\theta$ in the laboratory frame,
calculated relativistically (solid line) 
ad nonrelativistically (dashed line) for the incoming 
neutrino energy $E$= 150 MeV (left panel) and $E$= 300 MeV (right panel).
\label{f01}}
\end{figure}

In the nonrelativistic approximation the kinetic energy of the 
2N system in the 2N total momentum zero frame must be  
non-negative:
\begin{eqnarray} 
{ E} + M_d - 2 M_p - \sqrt{ M_l^2 + {{\bf k}^{\prime}}^{\, 2} \, }  
- \frac{ \left( {\bf k} - {\bf k}^{\prime} \, \right)^2 }{ 4 M_p } \ge 0  \, .
\label{nrlnn7}
\end{eqnarray}  
This condition yields now a forth degree equation:
\begin{eqnarray}
{k^\prime}^{\, 4} - 4 { E} \cos\theta {k^\prime}^{\, 3} \nonumber \\
+ \left( 4 { E}^2 \cos^2 \theta - 16 M_p^2 - 2 W \, \right) {k^\prime}^{\, 2} \nonumber \\ 
+ 4 W { E} \cos\theta {k^\prime} \, + \, W^2 - 16 M_p^2 M_l^2 = 0 \, ,
\label{nrlnn8}
\end{eqnarray}
with $ W \equiv 4 M_p \left( { E} + M_d - 2 M_p \right) - { E}^2 $.
One of the roots of Eq.~(\ref{nrlnn8}) is the nonrelativistic analogue 
of $ \left( k^\prime \right)_{max}^{rel} $, which we denote by 
$ \left( k^\prime \right)_{max}^{nrl} $. Its values are found numerically,
starting the search from $ \left( k^\prime \right)_{max}^{rel} $.
In Fig.~\ref{f01} we show a comparison of the kinematical constraints 
for $ \left(k^\prime \right)_{max}$, comparing $ \left( k^\prime \right)_{max}^{rel} $ 
and $ \left( k^\prime \right)_{max}^{nrl} $ for two initial electron 
neutrino energies, $E$= 150 and 300 MeV. Even at $E$= 300 MeV curves
representing two different results nearly overlap. 
The maximal difference is noticed for the backward angles,
but it does not reach 0.5~\%.

The last issue we want to discuss here concerns the phase space
factor. In Refs.~\cite{PRC63.034617,PRC86.035503} two different relativistic
forms are employed. Since in Ref.~\cite{PRC86.035503} the same variables 
are used as in our nonrelativistic framework, we focus on the effect
of the relativistic phase space factor from Ref.~\cite{PRC86.035503} 
on the inclusive cross sections.
With $ \omega \equiv E - \sqrt{ M_l^2 + {{\bf k}^{\prime}}^{\, 2} \, } \equiv E - E^\prime$
and $ {\bf Q} \equiv {\bf k} - {{\bf k}^{\prime}}$ we write
\begin{eqnarray}
\frac{d \sigma }{ d^3 {\bf k}^{\prime} } \sim 
\delta \left( \omega + M_d -
\sqrt{ M_p^2 + {\bf p}_1^{\, 2} \, }   
-
\sqrt{ M_p^2 + {\bf p}_2^{\, 2} \, } \, 
\right)
\,
\delta^3 \left( {\bf Q} -  {\bf p}_1 - {\bf p}_2 \, \right) \,
d^3{\bf p}_1 \, d^3{\bf p}_2  \, .
\label{psf1}
\end{eqnarray}  
By changing variables from the individual nucleons' momenta (${\bf p}_1$ and ${\bf p}_2$)
to the total (${\bf p}_{12}$) and relative one (${\bf p}$), 
%$ {\bf p}_{12} = {\bf p}_1 + {\bf p}_2 $,
%$ {\bf p} =  \left(  {\bf p}_1 - {\bf p}_2 \, \right) $
we rewrite (\ref{psf1}) as
\begin{eqnarray}
\frac{d \sigma }{ d^3 {\bf k}^{\prime} } \sim 
\delta \left( \omega + M_d -
\sqrt{ M_p^2 + \left( {\bf p} + \frac12 {\bf Q}  \right)^{\, 2} \, }   
-
\sqrt{ M_p^2 + \left( {\bf p} - \frac12 {\bf Q}  \right)^{\, 2} \, }   
\right) \, d^3{\bf p}  \, .
\label{psf2}
\end{eqnarray}  
Defining $ x \equiv \hat{\bf p} \cdot \hat{\bf Q} $ 
and $Q \equiv \mid {\bf Q} \mid$
we evaluate (\ref{psf2}) as
\begin{eqnarray}
\frac{d \sigma }{ d^3 {\bf k}^{\prime} } \sim 
\int\limits_{-1}^1 dx \, 
\frac{p^2}{ \left| 
\frac{p + Q x /2 }{ E_1} + \frac{p - Q x /2 }{ E_2} 
 \right| } \, ,
\label{psf4}
\end{eqnarray}  
where 
$E_{1,2} \equiv \sqrt{ M_p^2 + \left(  p \pm  Q x /2  \right)^{\, 2} \, }  $
and
$
p= \frac{ \left( M_d + \omega \right) \,  \sqrt{ -4 M_p^2 + ( M_d + \omega - Q ) ( M_d + \omega + Q )  } \, }
{ 2 \, \sqrt{  \left(M_d + \omega - Q x \right) \, \left( M_d + \omega + Q x \right) \, } }
$.

The corresponding nonrelativistic evaluation starts with 
\begin{eqnarray} 
\frac{d \sigma }{ d^3 {\bf k}^{\prime} } \sim 
\delta \left( \omega + M_d - 2 M_p - 
\frac { {\bf p}_1^{\, 2} \, }  { 2 M_p}  
-
\frac { {\bf p}_2^{\, 2} \, }  { 2 M_p} \, 
\right)
\,
\delta^3 \left( {\bf Q} -  {\bf p}_1 - {\bf p}_2 \, \right) \,
d^3{\bf p}_1 \, d^3{\bf p}_2 
\label{psf6}
\end{eqnarray} 
and the same change of variables yields  
after standard steps 
\begin{eqnarray}
\frac{d \sigma }{ d^3 {\bf k}^{\prime} } \sim 
\int\limits_{-1}^1 dx \,  \frac12 M_p \, p_{nrl} \, ,
\label{psf7}
\end{eqnarray}  
where $ p_{nrl} = \sqrt{ M_p \left( \omega + M_d - 2 M_p - \frac{{\bf Q}^2}{4 M_p} \,   \right) \, } $.

\begin{figure}
\includegraphics[width=7cm]{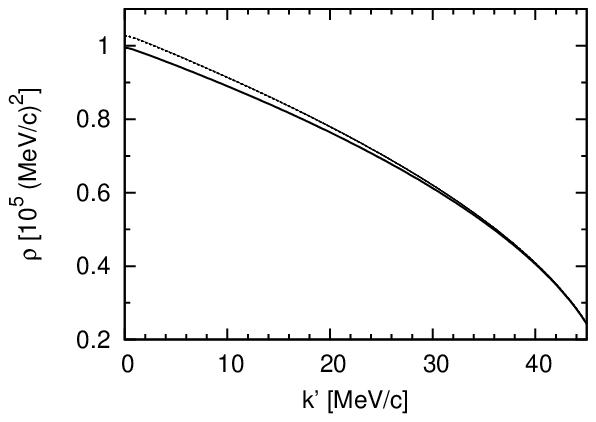}
\includegraphics[width=7cm]{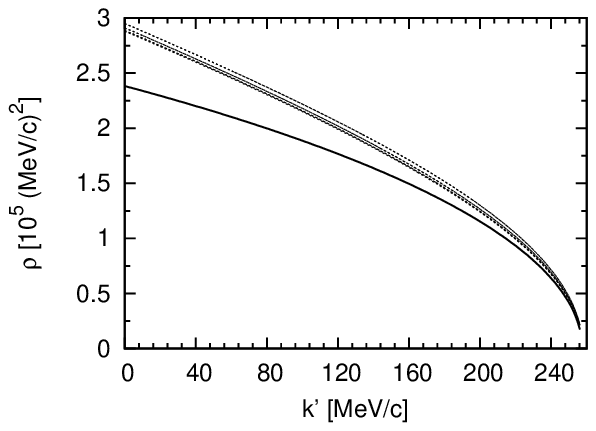}
\caption{The relativistic (group of dotted lines) and nonrelativistic (solid line) phase space 
factors compared for the initial neutrino energy $E$= 50 MeV (left panel)
and $E$= 300 MeV (right panel) for the lepton scattering angle $\theta = \pi/2$
as a function of the outgoing lepton momentum $k^\prime$.
The group of dotted curves represent results with different 
values of $x$ (see text). Note that these different 
predictions overlap for $E$= 50 MeV.
\label{FIG.PSF}}
\end{figure}

In Fig.~\ref{FIG.PSF} we compare the relativistic,
$ \rho_{rel} = 
\frac{p^2}{ \left| 
\frac{p + Q x /2 }{ E_1} + \frac{p - Q x /2 }{ E_2} 
 \right| } $,
and the nonrelativistic,
$ \rho_{nrl} =  \frac12 M_p \, p_{nrl} $,
phase space factors. 
For a fixed initial electron neutrino energy %$E$= 50 and $E$= 300 MeV
and just one lepton scattering angle $\theta = \pi/2$ we calculate 
the phase space factors as a function of the outgoing lepton momentum $k^\prime$.
The relativistic phase space factor depends not only on the magnitude
of the relative momentum $p$ but also on $x$, so we calculate $ \rho_{rel} $
for seven $x$ values: $-1$, $-2/3$, $-1/3$, $0$, $1/3$, $2/3$ and $1$ to check
how strong the dependence on $x$ is. 
For $E$= 50 MeV all upper curves representing relativistic results with 
different $x$ values essentially overlap, but for $E$= 300 MeV the spread 
due to the different $x$ values is clearly visible. The relativistic phase space factors are larger 
in the whole range of the $k^\prime$ momentum. The difference is particularly 
strong for small $k^\prime$ values and for $E$= 300 MeV it exceeds 20~\%.
For $E$= 50 MeV it is much smaller and reaches about 3~\%.

% FIG.1

% FIG. 2

\subsection{The 2N scattering states in coordinate and momentum spaces}
\label{section2.2}

The nuclear matrix element,
$\BA{\Psi_f \, {\bf P}_f \, m_{f} \, } \, 
j_{CC}^\lambda \, \KT{\Psi_i \, {\bf P}_i \, m_{i} \, } $,
involves the initial deuteron state 
and the 2N scattering state,
\begin{eqnarray} 
\BA{\Psi_f \, {\bf P}_f \, m_{f} \, } \, 
j_{CC}^\lambda \, \KT{\Psi_i \, {\bf P}_i \, m_{i} \, } \, = \,
^{(-)}\BA{ {\bf p} \  {\bf P}_f={\bf k} - {\bf k}^\prime \  m_{1} \, m_2 } \, 
j_{CC}^\lambda \, \KT{\phi_d \, {\bf P}_i=0 \  m_{d} \, } \nonumber \\
= \ \BA{ {\bf p} \  {\bf P}_f={\bf k} -{\bf k}^\prime \  m_{1} \, m_2 } \, 
\Big( \,  1 +  t (E_{2N} ) \, G_0^{2N} (E_{2N} ) \, \Big) \, 
j_{CC}^\lambda \, \KT{\phi_d \, {\bf P}_i=0 \  m_{d} \, } \nonumber \\
= \ \BA{ {\bf p} \  m_{1} \, m_2 } \, 
\Big( \,  1 +  t (E_{2N} ) \, G_0^{2N} (E_{2N} ) \, \Big) \, 
j_{CC}^\lambda \left( {\bf P}_f, {\bf P}_i \, \right)  \, \KT{\phi_d \, \  m_{d} \, } \, \equiv \, N^\lambda  \, ,
\label{nnn1}
\end{eqnarray}  
obtained, for a given NN potential $V$, 
from the $t$~matrix - solution of the Lippmann-Schwinger equation:
\begin{eqnarray} 
t (E_{2N}) = V +  t (E_{2N} ) \, G_0^{2N} (E_{2N} ) \, V \, ,
\label{t}
\end{eqnarray} 
where $ G_0^{2N} (E_{2N}) $ is the free 2N propagator 
and the kinetic energy of the relative motion in the 2N system 
in our nonrelativistic approximation,
$ E_{2N} = \frac { {\bf p}^{\, 2} \, }  { M} $,
 is given by Eq.~(\ref{nrlnn5}).
 
As already stated, we generate the deuteron wave function and 
solve Eq.~(\ref{t}) in momentum space, using 
the 2N partial wave states, $ \mid p (ls ) j m_j ; t m_t \rangle$. 
They carry information 
about the magnitude of the relative momentum ($p$),
the relative angular momentum ($l$), spin ($s$) and total angular
momentum ($j$) with its corresponding projection ($m_j$).
This set of quantum numbers is supplemented by the 2N 
isospin ($t$) and its projection ($m_t$).
In the present work we employ all partial wave states with $ j \le 4$.
Such calculations, closely corresponding to the ones presented in 
Ref.~\cite{PRC90.024001}, are fully sufficient 
for the antineutrino induced CC
reactions and the NC driven reactions,
where only short-range potentials act between the two outgoing nucleons.

\begin{figure}
\includegraphics[width=6cm]{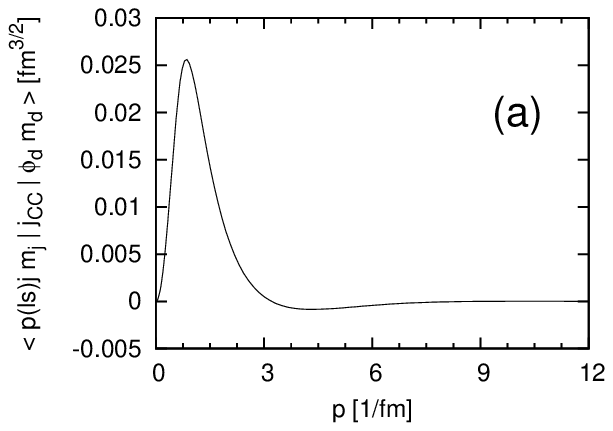}
\includegraphics[width=6cm]{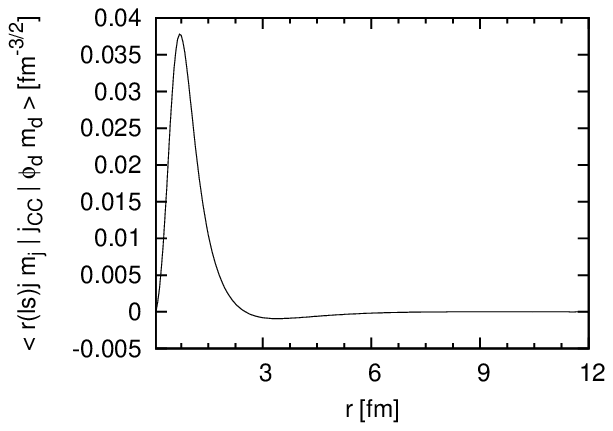}
\includegraphics[width=6cm]{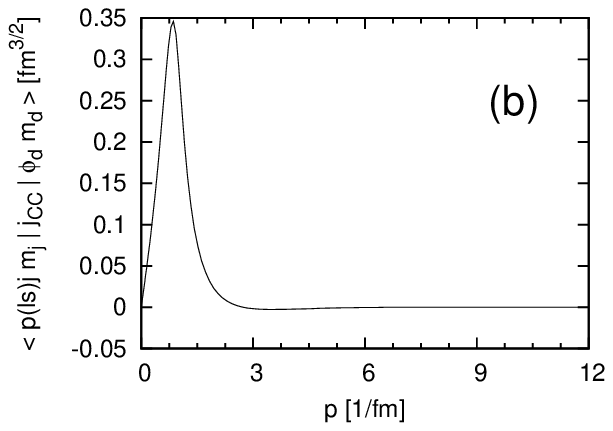}
\includegraphics[width=6cm]{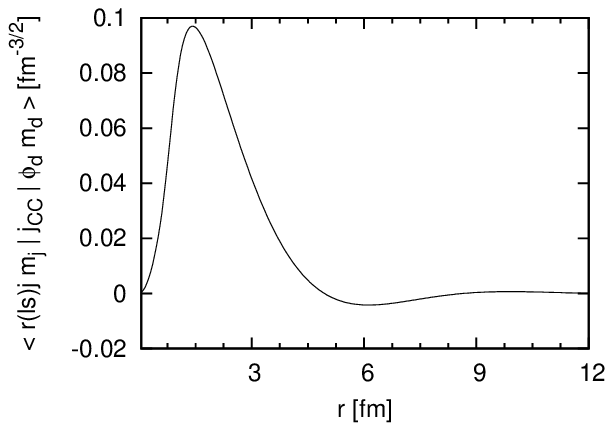}
\includegraphics[width=6cm]{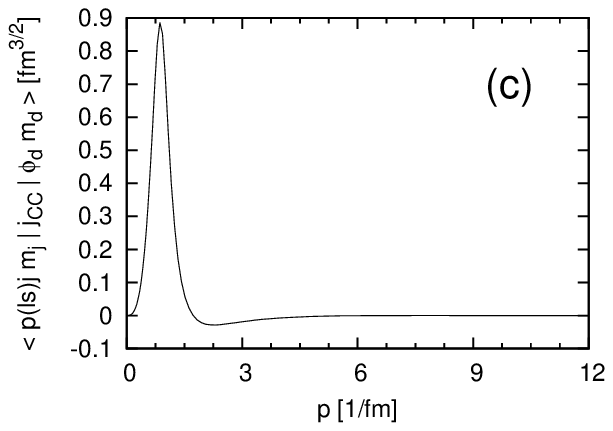}
\includegraphics[width=6cm]{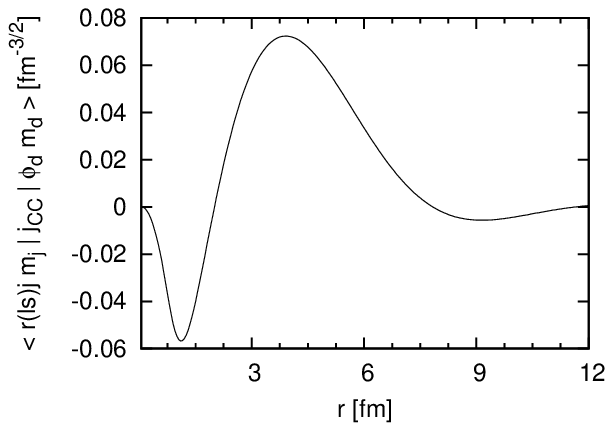}
\includegraphics[width=6cm]{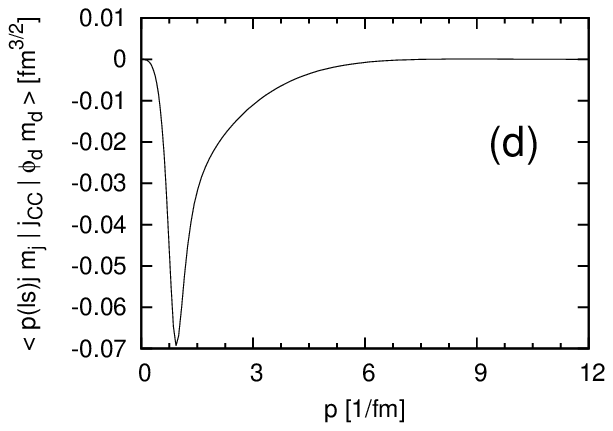}
\includegraphics[width=6cm]{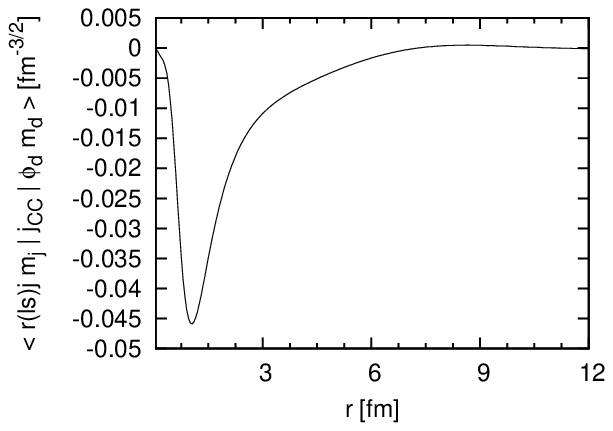}
\caption{Matrix elements of the single nucleon weak CC
operator in the partial wave basis of momentum space (left panel) 
and of coordinate space (right panel) for one selected lepton kinematics
of the
$\bar{\nu}_e + {^2{\rm H}} \rightarrow e^+ + n + n$
reaction: $E$= 300 MeV, $\theta$= 100$^\circ$, $E^\prime$= 172 MeV
($\omega$= 128 MeV, $Q$= 354 MeV, $E_{2N}$= 91 MeV).
The rows correspond to the different choices of
the projection of the total deuteron spin $m_d$, 
the component of the current operator $j_{CC}^\lambda$
and the final 2N channel. From top to bottom:
((a) $m_d$= 0, $j_{CC}^0$, $^1S_0$),
((b) $m_d$=-1, $j_{CC\, +1}$, $^3P_2$),
((c) $m_d$=+1, $j_{CC\, -1}$, $^3F_2$),
((d) $m_d$= 0, $j_{CC\, z}$, $^3F_2$).
Note that the factor $i^l$ is skipped in the Bessel transform.
\label{FIG.TR}}
\end{figure}

The neutrino induced CC reactions lead, however, in the final state 
to two protons,
which interact also by the long-range Coulomb potential.
The 2N scattering problem involving this interaction
is usually solved in coordinate space. We could follow the steps
outlined in Ref.~\cite{PRC86.035503}, but we wanted to take advantage of 
momentum space framework developed for the muon capture reaction.
That is why we decided to perform standard momentum space $t$~matrix calculations
for the short-range potential. Thus the proton-proton version of the AV18 
potential was supplemented by the sharply cut off Coulomb force $V_{RC}$,
whose matrix elements are given by the following integral
\begin{eqnarray}
\langle p^\prime (l^\prime s^\prime) j^\prime m_{j^\prime} ; t^\prime m_{t^\prime} 
\mid V_{RC} \mid p (l s)j m_j ; t m_t \rangle = \nonumber \\
\delta_{l l^\prime} \,
\delta_{s s^\prime} \,
\delta_{j j^\prime} \,
\delta_{m_j m_{j^\prime}} \,
\delta_{t t^\prime} \,
\delta_{m_t m_{t^\prime}} \,
\delta_{t 1} \,
\delta_{m_t 1} \,
8 \alpha \, \int\limits_0^{R_C} dr\, r \, 
j_{l} (p^\prime r ) \,
j_{l} (p r ) \, ,
\label{potc}
\end{eqnarray}
where $ j_{l} (p r ) $ is the spherical Bessel function
and $\alpha$ is the fine structure constant.
The value of the sharp cut-off was taken to be $R_C$= 40 fm.
This approach is fully justified by the observation
that the current matrix elements Bessel transformed to coordinate space,
\begin{eqnarray}
\langle r (ls)j m_j \mid j_{CC}^\lambda \mid \phi_d m_d \rangle \, = \, 
\frac{2}{\pi} \, i^l \, \int\limits_0^\infty  dk\, {k}^{2} \, j_{l} (k r ) \,
\langle k (ls)j m_j \mid j_{CC}^\lambda \mid \phi_d m_d \rangle \, ,
\label{jinrspace} 
\end{eqnarray} 
become negligible for $r \ge $ 30~fm. 
This is illustrated in Fig.~\ref{FIG.TR} for one (essentially arbitrary)
lepton kinematics and few choices of discrete quantum numbers.

\begin{figure}
\includegraphics[width=6cm]{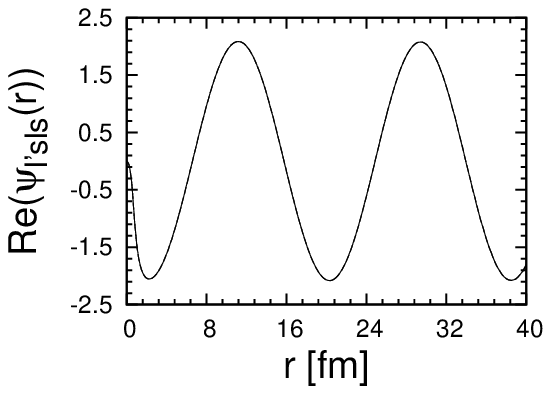}
\includegraphics[width=6cm]{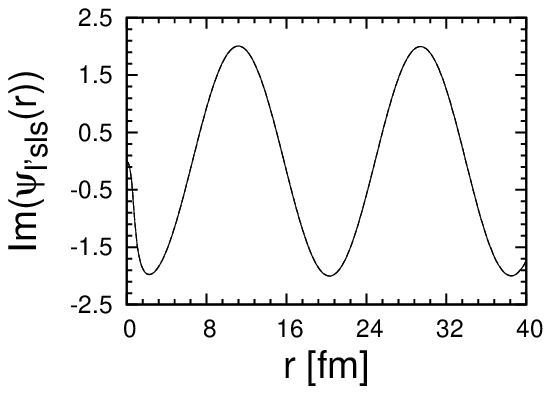}
\caption{The scattering radial wave function $\psi_{l's\, ls}(r)$ 
obtained directly in coordinate space (solid line) 
compared to the one generated from the solution of the Lippmann-Schwinger
equation obtained in momentum space (dashed line) for the internal 
2N energy $E_{2N}$= 5~MeV for the uncoupled $^1S_0$ channel.
The wave functions are calculated with the strong proton-proton
potential augmented by a sharply cut off Coulomb potential 
with the cut-off value $R_C$= 40~fm. The real (imaginary) parts of the 
wave functions are displayed in the left (right) panel. The solid and dashed lines
fully overlap.
\label{FIG.CWF.1}}
\end{figure}

\begin{figure}
\includegraphics[width=6cm]{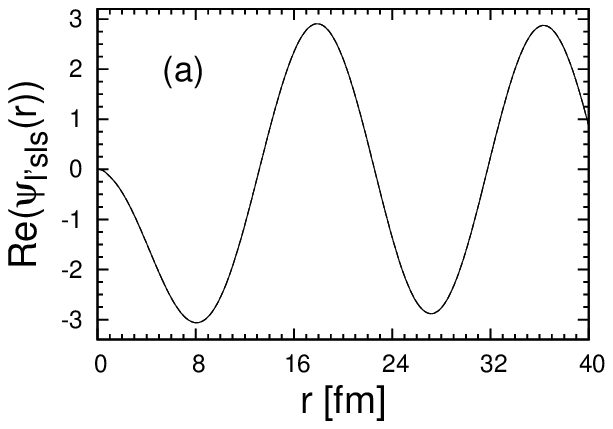}
\includegraphics[width=6cm]{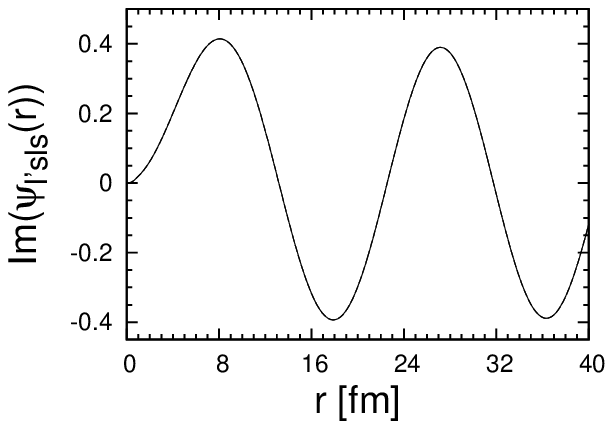}
\includegraphics[width=6cm]{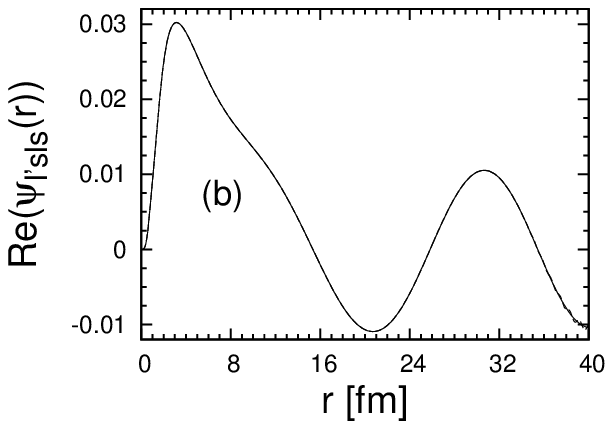}
\includegraphics[width=6cm]{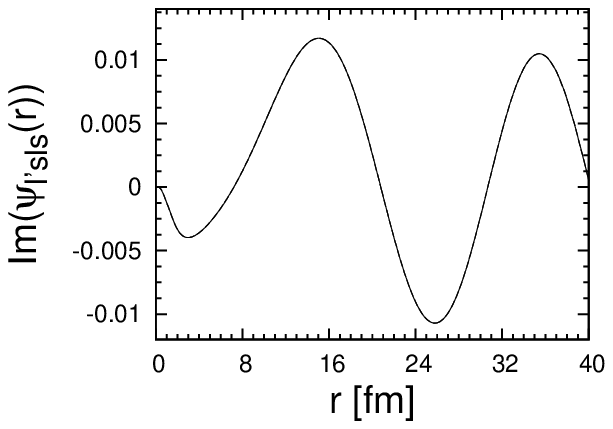}
\includegraphics[width=6cm]{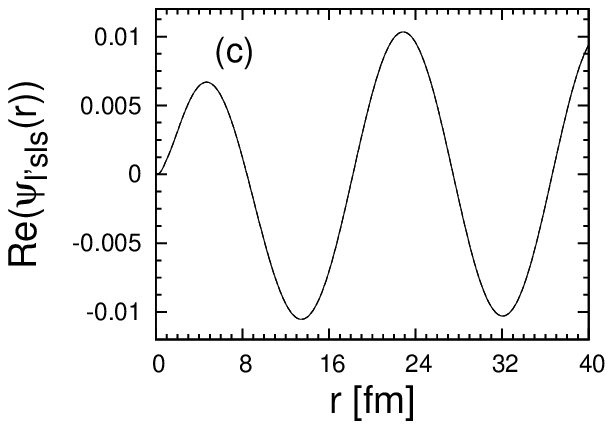}
\includegraphics[width=6cm]{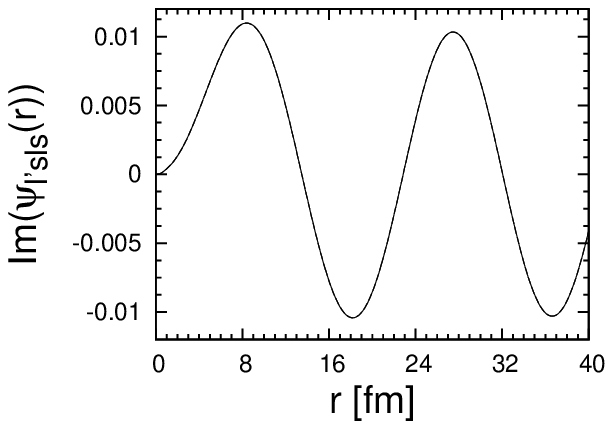}
\includegraphics[width=6cm]{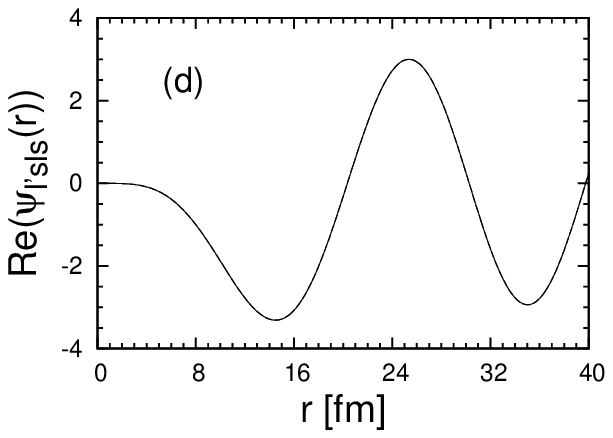}
\includegraphics[width=6cm]{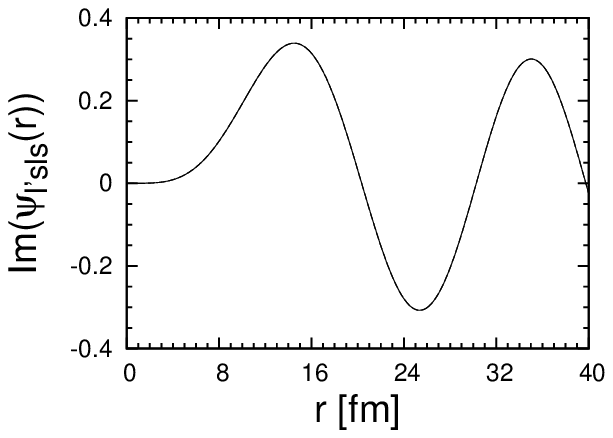}
\caption{The same as in Fig.~\ref{FIG.CWF.1} but for the coupled 
$^3P_2- ^3F_2$ channels. The real (imaginary) parts of the 
wave functions are displayed in the left (right) panels 
and the rows from top to bottom correspond to different $l' l$ pairs: 
((a) $^3P_2- ^3P_2$),  
((b) $^3F_2- ^3P_2$),  
((c) $^3P_2- ^3F_2$) and
((d) $^3F_2- ^3F_2$).
\label{FIG.CWF.2}}
\end{figure}

\begin{figure}
\includegraphics[width=6cm]{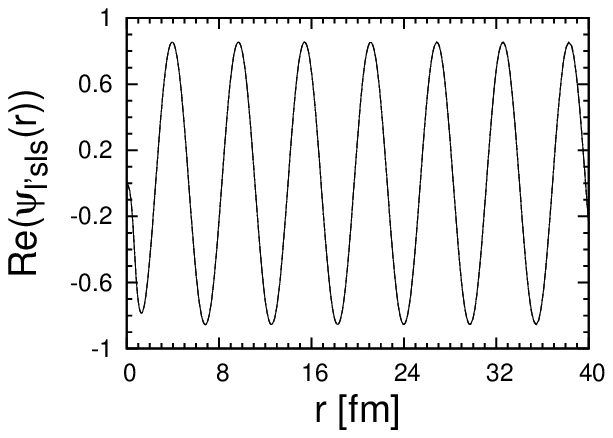}
\includegraphics[width=6cm]{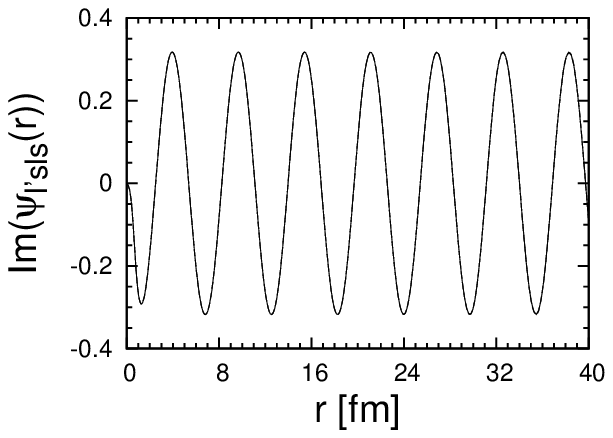}
\caption{The same as in Fig.~\ref{FIG.CWF.1} but for the internal 
2N energy $E_{2N}$= 50~MeV.
\label{FIG.CWF.3}}
\end{figure}

\begin{figure}
\includegraphics[width=6cm]{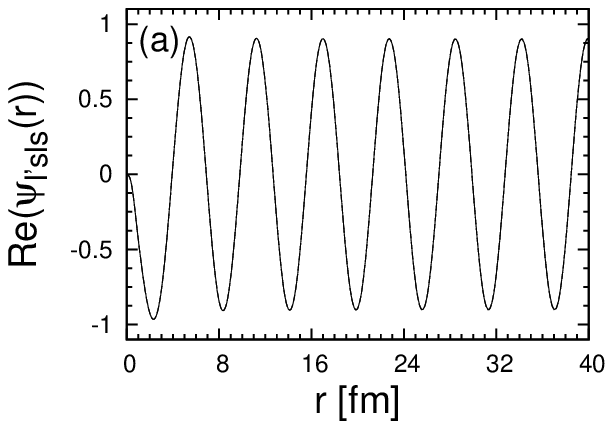}
\includegraphics[width=6cm]{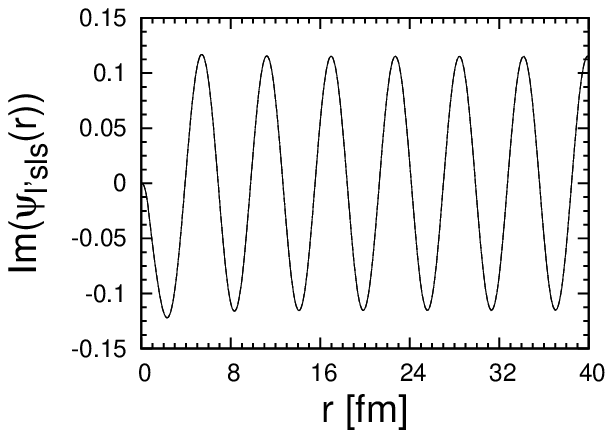}
\includegraphics[width=6cm]{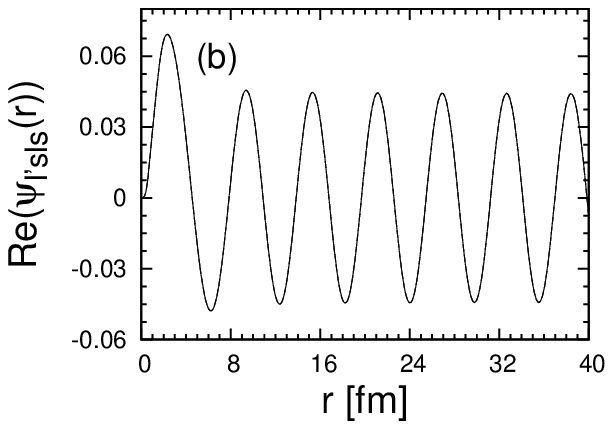}
\includegraphics[width=6cm]{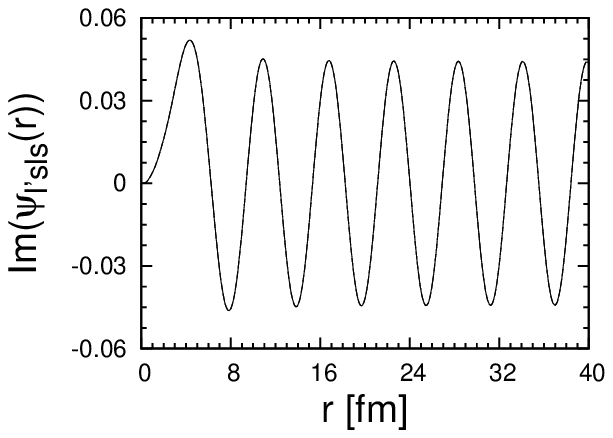}
\includegraphics[width=6cm]{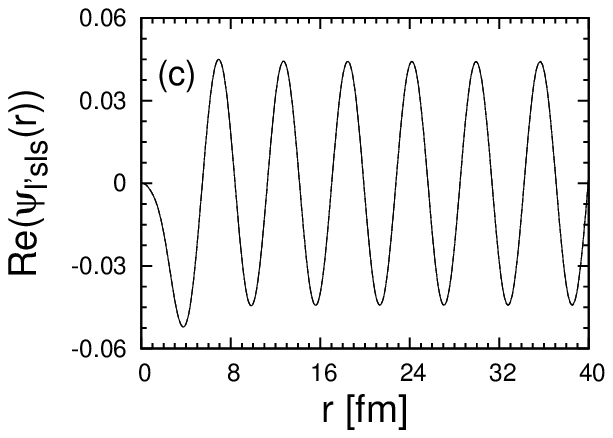}
\includegraphics[width=6cm]{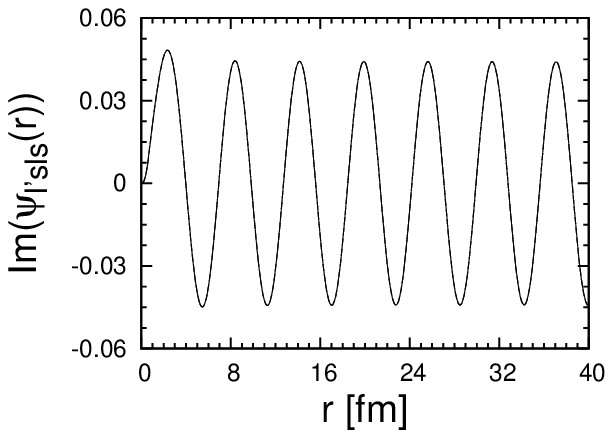}
\includegraphics[width=6cm]{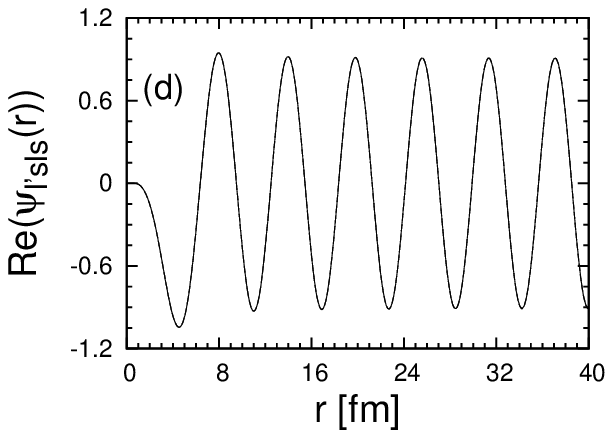}
\includegraphics[width=6cm]{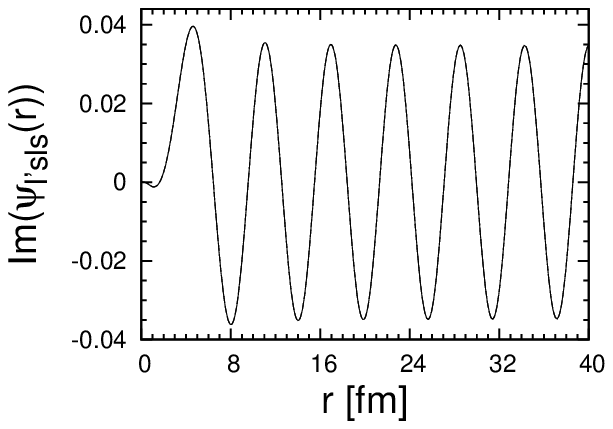}
\caption{The same as in Fig.~\ref{FIG.CWF.2} but for the internal 
2N energy $E_{2N}$= 50~MeV.
\label{FIG.CWF.4}}
\end{figure}

Additionally we checked that our momentum space generated 2N scattering 
states are fully equivalent to the radial wave functions calculated directly
in coordinate space, using the collocation method from 
Refs.~\cite{Lanczos38,deBoor73,Schellingerhout95}.
To this end we employed the well-known 
formula (see for example \cite{PRC86.035503} and references therein), which 
using our normalization of states, reads
\begin{eqnarray}
\psi_{l^\prime s\, l s }^j (r) = 
\delta_{l^\prime \, l } \, j_l (p r )  + 
i^{l - l^\prime} \, M \int\limits_0^\infty \frac{ dk\, {k}^{2} \, j_{l^\prime} (k r ) }
{p^2 - k^2 + i \epsilon} \, \langle k (l^\prime s ) j \mid t (E_{2N}) \mid p (l s ) j \rangle \, .
\label{checkwf} 
\end{eqnarray} 
This is exemplified in Figs.\ref{FIG.CWF.1}--\ref{FIG.CWF.4}, for two 
$E_{2N}$ energies (5 and 50 MeV) and two partial wave cases 
($^1S_0$ and $^3P_2- ^3F_2$). In all cases we get a perfect agreement 
in the whole range of $r$ values, from $r=0$ to $r$= 40~fm.

% FIG. 3

% FIG. 4

% FIG. 5

% FIG. 6

% FIG. 7

\subsection{The cross section}
\label{section2.3}
 
Having discussed all the elements of our formalism, we can give
the main formula for the cross section, consistent with the momentum 
space formalism presented in the previous subsections. We do it for the 
$\nu_e + {^2{\rm H}} \rightarrow e^- + p + p$ reaction
and discuss later some differences if other reactions are considered.
With our normalization of the $N^\lambda$ matrix elements we start with \cite{bjodrell}

\begin{eqnarray}
d \sigma & = & \frac1{ \mid {\bf v}_1 - {\bf v}_2 \mid } \, \frac1{2 E} \, 
\left( L_\alpha  \right)^* \, L_\beta \, 
\left( N^\alpha  \right)^* \, N^\beta \,  \frac{G_F^2 \, \cos^2\theta_C }{2} \nonumber \\
& \times & F(Z, k^\prime) \, \frac{d^3{\bf k}^\prime}{2 E^\prime } \, 
\frac{d^3{\bf p}_1}{ \left( 2 \pi  \right)^3 } \, 
\frac{d^3{\bf p}_2}{ \left( 2 \pi  \right)^3 } \, 
 \left( 2 \pi  \right)^4 \, \delta^4  \left( P^\prime - P \right) \, {\cal S} \, ,
\label{dsigma1}
\end{eqnarray}
where in the laboratory frame the relative velocity of the projectile
and target $ \mid {\bf v}_1 - {\bf v}_2 \mid $ is equal $c\, (= 1)$ and the ${\cal S}$ factor
is needed when in the final state identical particles appear.
The Fermi function $F(Z, k^\prime)$ \cite{Fermi34} is introduced to account for the Coulomb modification
of the final lepton wave functions by the two protons in the final state 
and is not needed otherwise.
(Note that in $L_\alpha$ we use the following normalization for the Dirac spinors:
$ \bar{u} \, u = 2 M_l $ and $ \bar{v} \, v = -2 M_l $.) 
Defining 
\begin{eqnarray}
\widetilde{L_{\alpha\beta}} = \sum\limits_{m_\nu} \,  \sum\limits_{m_{l^-}} \, \left( L_\alpha  \right)^* \, L_\beta 
\label{dsigma2} \, ,
\end{eqnarray}
taking all factors into account and evaluating the phase space factor 
in terms of the relative momentum, we arrive at the following 
expression for the total cross section 
\begin{eqnarray}
\sigma_{tot} & = & \frac{ G_F^2 \cos^2\theta_C} { 2 \left( 2 \pi \right)^2 } \, \frac{1}{ 4 E } \, 
\int\limits_0^{2\pi} d \phi \, \int\limits_0^{\pi} d \theta \, \sin\theta \, 
\int\limits_{M_l}^{{E^\prime}_{max}} d E^\prime \, k^\prime \, \frac12 \, M_p \, p_{nrl} \nonumber \\
& \times & F(Z, k^\prime) \, \int\limits_0^{2\pi} d \phi_p \, \int\limits_0^{\pi} d \theta_p \, \sin\theta_p \,
\frac23 \, \sum\limits_{m_1, m_2} \sum\limits_{m_d} \widetilde{L_{\alpha\beta}} \, 
\left( N^\alpha  \right)^* \, N^\beta \, ,
\label{dsigma3}
\end{eqnarray}  
where the $\theta$ and $\phi$ angles describe the direction
of the outgoing lepton in the laboratory frame. 
Note that for the total unpolarized cross section considered here 
the integral over the azimuthal angle $\phi$ 
can be replaced by the factor $2 \pi$.

The contraction $ \mid T \mid^2 \equiv \widetilde{L_{\alpha\beta}} \, \left( N^\alpha  \right)^* \, N^\beta $ 
can be written in terms of the $V_{ij}$ functions stemming from the lepton arm 
and the products of the nuclear matrix elements $  N^\alpha $. For the
latter we use the spherical components and obtain 
\begin{eqnarray}
\mid T \mid^2 & = &
V_{00} \mid N^0 \mid^2 \, + \, 
V_{MM} \mid N_{-1} \mid^2 \, + \, 
V_{PP} \mid N_{+1} \mid^2 \, \nonumber \\
& + &  V_{ZZ} \mid N_{z} \mid^2 \, + \, 
V_{Z0} N_{z} \, \left( N^0 \right)^* \, + \, 
V_{0Z} N^{0} \, \left( N_z \right)^* \, ,
\label{Tsquared}
\end{eqnarray}
where for the neutrino induced reactions 
\begin{eqnarray}
V_{00} & = & 8 \, \left(  {\bf k}^{\, \prime} \cdot {\bf k} + E \, E^\prime \, \right) \, , \nonumber \\
V_{MM} & = & 8 \, \left( E + k_z \, \right) \, \left( E^\prime - k^\prime_z \, \right) \, , \nonumber \\
V_{PP} & = & 8 \, \left( E - k_z \, \right) \, \left( E^\prime + k^\prime_z \, \right) \, , \nonumber \\
V_{ZZ} & = & 8 \, \left(  -{\bf k}^{\, \prime} \cdot {\bf k} + E \, E^\prime \, \right) \, , \nonumber \\
V_{Z0} & = & -8 \, \left( E \, k^\prime_z  + E^\prime \, k_z \, \right) \, + 8 i \, \left( k^\prime_y \, k_x \, - \, k^\prime_x \, k_y \, \right) \, , \nonumber \\
V_{0Z} & = & -8 \, \left( E \, k^\prime_z  + E^\prime \, k_z \, \right) \, + 8 i \, \left( k^\prime_x \, k_y \, - \, k^\prime_y \, k_x \, \right) \, .
\label{V-fun}
\end{eqnarray}

The corresponding $\bar{V}_{ij}$ functions for
the antineutrino induced reactions are given as:
\begin{eqnarray}
\bar{V}_{00} & = & V_{00}  \, , \nonumber \\
\bar{V}_{MM} & = & V_{PP}  \, , \nonumber \\
\bar{V}_{PP} & = & V_{MM}  \, , \nonumber \\
\bar{V}_{ZZ} & = & V_{ZZ}  \, , \nonumber \\
\bar{V}_{Z0} & = & V_{0Z}  \, , \nonumber \\
\bar{V}_{0Z} & = & V_{Z0}  \, .
\label{BARV-fun}
\end{eqnarray}

In the following we assume the system of coordinates, where $ {\bf Q} \equiv {\bf k}  -  {\bf k}^{\, \prime} \parallel {\hat z}$
and ${\hat y} = \frac{ {\bf k} \times  {\bf k}^{\, \prime} } {\mid  {\bf k} \times  {\bf k}^{\, \prime} \mid }$, so 
\begin{eqnarray}
k^\prime_x & = & k_x  =  \mid  {\bf k} \mid \mid  {\bf k}^{\, \prime} \mid \sin \theta / \mid  {\bf Q} \mid \, , \nonumber \\
k^\prime_y & = & k_y  =  0 \, , \nonumber \\
k_z & = & \mid  {\bf k} \mid \left( \mid  {\bf k} \mid  - \mid  {\bf k}^{\, \prime} \mid \cos \theta \, \right)  / \mid  {\bf Q} \mid \, , \nonumber \\
k^\prime_z & = & \mid  {\bf k}^{\, \prime}  \mid \left( -\mid  {\bf k}^{\, \prime}  \mid  + \mid  {\bf k} \mid \cos \theta \, \right)  / \mid  {\bf Q} \mid \, , \nonumber \\
\mid  {\bf Q} \mid & = & \sqrt{ {\bf k}^{\, 2} +  {{\bf k}^{\, \prime \, 2 }} - 2 \mid  {\bf k} \mid  \mid  {\bf k}^{\, \prime} \mid \cos \theta \, } \, .
\label{kinemtheta}
\end{eqnarray}
As a consequence, we get further simplifications:
\begin{eqnarray}
\bar{V}_{0Z} = \bar{V}_{Z0} = V_{0Z} = V_{Z0}  \, .
\label{BARV-fun2}
\end{eqnarray}
The main reason to use the spherical components of the current operator 
and the system of coordinates defined above is that we get in this case
the simplest relations between the total spin magnetic quantum numbers $m_d$ and $m_j$ 
for matrix elements
$\langle p (ls)j m_j \mid j_{CC}^\lambda \mid \phi_d m_d \rangle $
in the partial wave representation:
\begin{eqnarray}
\langle p (ls)j m_j \mid j_{CC}^0 \mid \phi_d m_d \rangle \propto \delta_{m_j, m_d} \, ,
\\ \nonumber
\langle p (ls)j m_j \mid j_{CC, z} \mid \phi_d m_d \rangle \propto \delta_{m_j, m_d} \, ,
\\ \nonumber
\langle p (ls)j m_j \mid j_{CC, \pm 1} \mid \phi_d m_d \rangle \propto \delta_{m_j \pm 1, m_d} \, .
\label{relation-m}
\end{eqnarray}

For the neutral current (NC) driven processes Eq.~(\ref{dsigma3}) has to be modified.
The Fermi function $F(Z, k^\prime) $ and 
$ \cos^2\theta_C$ are replaced by $1$, but most importantly the weak 
CC operator $j_{CC}^\lambda$ is replaced by the corresponding 
NC operator $j_{NC}^\lambda$. Its construction is described in detail 
in Refs.~\cite{PRC63.034617,PRC86.035503} and we follow Ref.~\cite{PRC86.035503} 
for the choice of the nucleon form factors. Since we employ only the single nucleon current,
we use the given prescription for the proton and neutron NC
operators and, using the isospin formalism, define the current of nucleon $i$ as
\begin{eqnarray}
j_{NC}(i) = 
\frac12 \left( 1 + {\tau}_3 (i) \right) j_{NC}^p  \, + \, 
\frac12 \left( 1 - {\tau}_3 (i) \right) j_{NC}^n 
\label{jNC}
\end{eqnarray}
in full analogy to the electromagnetic single nucleon current.
Of course, also in this case the relations (\ref{relation-m}) remain true.

\subsection{The 3N matrix elements}
\label{section2.4}
 
We treat the $^3$He ($^3$H) disintegration process analogously to the 2N 
reactions.
The 3N Hamiltonian $H$ comprises the 3N kinetic energy ($H_0$),
two-body subsystem potential energies ($V_{12}$, $V_{23}$ and $V_{31}$) as well as the 
three-body potential energy ($V_{123}$). The latter quantity is usually called a 3N force (3NF)
and is decomposed into three terms
\begin{eqnarray}
V_{123} = V^{(1)} +  V^{(2)} +  V^{(3)} \, , 
\label{e3nf_split}
\end{eqnarray}
where $V^{(i)}$ is symmetric under exchange of nucleons
$j$ and $k$ ($i, j, k = 1, 2, 3$, $i \ne j \ne k \ne i$). 
The 3N bound state wave function is calculated using the method
described in Ref.~\cite{Nogga.1997}. The Faddeev equation for the Faddeev 
component $\mid \psi \rangle$ reads
\begin{eqnarray}
\mid \psi \rangle  = G_0 t_{23}  P \mid  \psi  \rangle + (1+ G_0 t_{23} ) G_0 V^{(1)} (1+P) \mid \psi \rangle \;.
\label{eq.bs}
\end{eqnarray}
Here $G_0\equiv 1/\left( \, E-H_0 \, \right) $ is the free 3N propagator and
$P \equiv P_{12} P_{23} + P_{13} P_{23}$ 
is the permutation operator built from transpositions
$P_{ij}$, which interchange nucleons $i$ and $j$.
Note that the two-body $t$-operator $t_{23}$ acts now in the 3N space.
The full wave function $\mid \Psi \rangle $ is easily obtained from the Faddeev component as
\begin{eqnarray}
 \mid \Psi \rangle = ( 1 + P ) \mid \psi \rangle \, .
\end{eqnarray}

The 3N current operator $j^{\mu}_{3N} $ contains the
single-nucleon, 2N and, in principle, also the 3N contribution. Therefore 
we write:
\begin{eqnarray}
j^{\mu}_{3N} = j^{\mu}_{1} + j^{\mu}_{2} + j^{\mu}_{3} 
             + j^{\mu}_{12} + j^{\mu}_{23} + j^{\mu}_{31} + j^{\mu}_{123} \, ,
\end{eqnarray}
where the 3N part can be split into three components (just like the 3NF),
$ j^{\mu}_{123} = j^{\mu\, 1} +   j^{\mu\, 2} + j^{\mu\, 3} $.
Thus we can decompose the 3N current operator into three 
parts, $j^{\mu} (i) $ ($i=1,2,3$),  which possess the same symmetry properties as 
$ V^{(i)} $:
\begin{eqnarray}
j^{\mu}_{3N} =   j^{\mu} (1) + j^{\mu} (2) + j^{\mu} (3) \, ,
\end{eqnarray}
where for example $ j^{\mu} (1) \equiv  j^{\mu}_{1} + j^{\mu}_{23} +  j^{\mu\, 1} $.

With all these ingredients, we construct the matrix elements 
for the nucleon-deuteron (Nd)
\begin{equation}
\label{eq52B}
 N^{\mu}_{Nd} = \langle \Psi^{(-)}_{Nd} \mid j^{\mu}_{3N} \mid \Psi\rangle
\end{equation}
and the 3N break-up channel
\begin{equation}
\label{eq52C}
 N^{\mu}_{3N} = \langle \Psi^{(-)}_{3N} \mid j^{\mu}_{3N} \mid \Psi\rangle \, ,
\end{equation}
with the corresponding channel scattering states.
To this end first we solve a Faddeev-type equation for an auxiliary state $ \mid U^\mu \,  \rangle $ \cite{raport2005},
\begin{eqnarray}
\label{U3n}
\mid U^\mu \,  \rangle & = & \Big( \; t_{23} G_0 \; +  \;\frac12 \left( \; 1 + P \: \right) \: V^{(1)} G_0 \; \left( \; 1 + t_{23} G_0 \: \right)   \Big)
\, ( 1 + P ) j^{\mu} (1) \mid \Psi \, \rangle  \nonumber \\
 & + & \;  \Big( \; t_{23}  G_0 P \; +  \;\frac12 \left( \; 1 + P \: \right) \: V^{(1)} G_0 \; \left( \; 1 + t_{23}  G_0 \: \right) P   \Big) \; \mid U^\mu \, \rangle  \,,
\end{eqnarray}
which depends on the component of the current operator and two kinematical quantities,
but is independent of the final state kinematics. The two
kinematical quantities are the 3N internal energy $E_{c.m.}$
and the magnitude of the three-momentum transferred to the 3N system, $\mid \bf Q \mid $.
In practice we use the density operator $ \rho \equiv j^0_{3N} $
as well as the spherical components of the current operator 
\begin{eqnarray}
j_{+1} \equiv -\frac1{\sqrt{2}} \, \left( j^1_{3N}  + i j^2_{3N}  \, \right) 
\, \equiv \,  -\frac1{\sqrt{2}} \, \left( j_{x\, 3N}  + i j_{y\, 3N}  \, \right) 
\, , \nonumber \\
j_{-1} \equiv  \frac1{\sqrt{2}} \, \left( j^1_{3N}  - i j^2_{3N}  \, \right) 
\, \equiv \,   \frac1{\sqrt{2}} \, \left( j_{x\, 3N}  - i j_{y\, 3N}  \, \right) 
\end{eqnarray}
and choose $\bf Q$ parallel to the $z$-axis. As in the 2N case, this yields the simplest 
relations between the total 3N angular momentum projections of the initial 
and final nuclear systems.

The matrix elements $N^{\mu}_{Nd}$ and $N^{\mu}_{3N}$ for arbitrary exclusive kinematics 
are then obtained by simple quadratures:
\begin{equation}
N^{\mu}_{Nd} = \langle \phi_{Nd} \mid (1+P) j^{\mu} (1) \mid \Psi \rangle \, + \, \langle \phi_{Nd} \mid P \mid U^{\mu} \rangle  \, ,
\end{equation}
\begin{eqnarray}
N^{\mu}_{3N} & = & \langle \phi_{3N} \mid (1+P) j^{\mu} (1) \mid \Psi \rangle \, + \, 
        \langle \phi_{3N} \mid t_{23}  G_0 \, \left( 1 + P \right) \, j^{\mu} (1) \,  \mid \Psi \rangle  \nonumber \\
   & + &   \langle \phi_{3N} \mid P \mid U^{\mu} \rangle \, + \,  \langle \phi_{3N} \mid t_{23}  G_0 P \mid U^{\mu} \rangle \,,
\end{eqnarray}
where $\mid \phi_{Nd} \rangle$ is a product of the internal deuteron state 
and the state describing the free relative motion of the third nucleon 
with respect to the deuteron
and $\mid \phi_{3N} \rangle$ is a state 
(antisymmetrized in the $(2,3)$ subsystem)
representing the free motion of the three outgoing nucleons.
Exclusive observables can be further integrated
over suitable phase space domains to arrive at the 
semi-exclusive or inclusive observables.

Inclusive observables can be, however, also computed in a different way,
without any resort to explicit final state kinematics~\cite{incRAB,raport2005}. 
In inclusive calculations, where only the final energy $E$
of the nuclear system is known, one encounters the so-called "response
functions", which are defined through the following integral
\begin{eqnarray}
R_{AB}^{inc}  = \sum\limits_{m_i,m_f} \int df \,
\delta \left( E - E_f  \right) \,
  \langle \Psi^{(-)}_{f} \mid j^{A}_{3N} \mid \Psi\rangle \,
\left( \langle \Psi^{(-)}_{f} \mid j^{B}_{3N} \mid \Psi\rangle \right)^* \, ,
\label{RAB}
\end{eqnarray}
and depend, in general, on two components of the nuclear current operator,
$A$ and $B$. In Eq.~(\ref{RAB}) $m_i$ and $m_f$ represent the whole sets 
of the initial and final spin magnetic quantum numbers, respectively, while 
the $df$ integral denotes the sum and the integration over all final
3N states with the energy $E$. Using closure, Eq.~(\ref{RAB}) can be rewritten
as
\begin{eqnarray}
R_{AB}^{inc}  =  \sum\limits_{m_i} \int df \,
  \langle \Psi \mid \left( j^{B}_{3N} \right)^\dagger \,
\delta \left( E - H  \right) \,
 j^{A}_{3N} \mid \Psi\rangle \, ,
\label{RAB2}
\end{eqnarray}
where $H$ is again the full 3N Hamiltonian and the 3N bound state does 
not contribute to the $df$ integration for $ E >0 $. Within the Faddeev
scheme  \cite{incRAB,raport2005}, $ R_{AB}^{inc} $ can be expressed in terms 
of some auxiliary states as
\begin{eqnarray}
R_{AB}^{inc}  =  
\frac1{ 2 \pi i } \, 
\sum\limits_{m_i} 
\Big(
{\langle \Psi \mid \left( j^{A}_{3N} \right)^\dagger \mid \Psi^B \rangle}^* 
 - 
 \langle \Psi \mid \left( j^{B}_{3N} \right)^\dagger \mid \Psi^A \rangle \, 
\Big) \nonumber \\
 = 
\frac3{ 2 \pi i } \, 
\sum\limits_{m_i} 
\Big(
{\langle \Psi \mid \left( j^{A} (1) \right)^\dagger G_0 ( 1 + P ) \mid V^B \rangle}^* 
 - 
 \langle \Psi \mid \left( j^{B} (1) \right)^\dagger G_0 ( 1 + P ) \mid V^A \rangle \, 
\Big)  ,
\label{RAB4}
\end{eqnarray}
where in turn ($C=A,B$) 
\begin{eqnarray}
\mid \Psi^C\rangle = G_0 ( 1 + P )  \mid V^C \rangle \, .
\end{eqnarray}
The state $  \mid V^C \rangle $ obeys the Faddeev-type equation 
\begin{eqnarray}
\label{V3n}
\mid V^C \,  \rangle & = &
\left( 1 + t_{23} G_0 \; \right) \, j^{C} (1) \mid \Psi \, \rangle  \nonumber \\
& + &  \Big( \; 
t G_0 P \; +  \; 
\left( \; 1 + t_{23} G_0 \: \right) 
V^{(1)} G_0 \; 
\left( \; 1 + P \: \right) \: 
  \Big) \; \mid V^C \, \rangle  \, ,
\end{eqnarray}
with the same integral kernel as in the treatment of 3N scattering \cite{Twith3NF}.
Interestingly, the relation between
the auxiliary states defined in Eqs.~(\ref{U3n}) and (\ref{V3n}) is
very simple when the 3N force is neglected:
\begin{eqnarray}
\label{UV3n}
 \mid V^C \, \rangle =  j^{C} (1) \mid \Psi \, \rangle \, + \,  \mid U^C \, \rangle \, .
\end{eqnarray}
The fact that $R_{AB}^{inc}$ can be obtained by direct integrations or employing the
method described above is used by us to test the numerical performance. 

All the 3N Faddeev-type equations are solved by iterations in the mo\-men\-tum-spa\-ce basis
\begin{equation}
\label{alpha}
\mid p\, q \, \alpha \rangle \equiv \mid p \, q\, (l s) j \, \left(\lambda \, \frac{1}{2} \right) I (jI)J m_J \,
\left( t \frac{1}{2} \right) T m_T  \rangle,
\end{equation}
which is an extension of the $(2,3)$ subsystem basis $\mid p \alpha_2 \rangle $.
Here $q$ is the magnitude of the Jacobi momentum, which describes the 
motion of the spectator nucleon $1$ with respect to the center of mass of the $(2,3)$ subsystem.
Consequently, the orbital angular momentum $\lambda$ of the spectator nucleon 
and its spin $\frac{1}{2}$ couple to the total spectator angular momentum $I$.
The total angular momentum of the subsystem ($j$) and the total angular 
momentum of nucleon $1$ ($I$) couple eventually to the
total angular momentum of the 3N system $J$ and its projection $m_J$. A corresponding
coupling is introduced in the isospin space, where the $(2,3)$ total subsystem isospin ($t$)
together with the isospin of nucleon $1$ builds the total 3N isospin $T$ with the projection $m_T$.
In practise, the calculations are restricted to a finite set of $ \mid p \, \alpha_2 \rangle$ 
and  $ \mid p \, \alpha \rangle$ states, which fulfill the condition $ j \le j_{max}$ and 
$J \le J_{max}$. For the studied here 
low 3N internal energies and momentum transfers 
convergence is achieved already with $j_{max}=3$ and $J_{max}=\frac{15}2$.

In the following sections we describe our results, which are obtained, like 
in Ref.~\cite{PRC86.035503}, without considering radiative corrections. 
Information about possible modifications of the results 
due to these effects are discussed in Ref.~\cite{Baroni17} and
references cited therein. 

\section{Results for neutrino scattering on $^2$H}
\label{section3}

Although on the way to calculate the total cross sections we evaluate 
the necessary integrands - the differential cross sections - we show here only 
the inclusive observables. Whenever it is possible, we compare our 
momentum-space results with the predictions from 
Shen {\it et al.} ~\cite{PRC86.035503}, which are also 
based on the traditional dynamical input. Since our calculations are strictly
nonrelativistic, we restrict ourselves
to the neutrino energies up to 250 MeV.

\begin{figure}
\includegraphics[width=6cm]{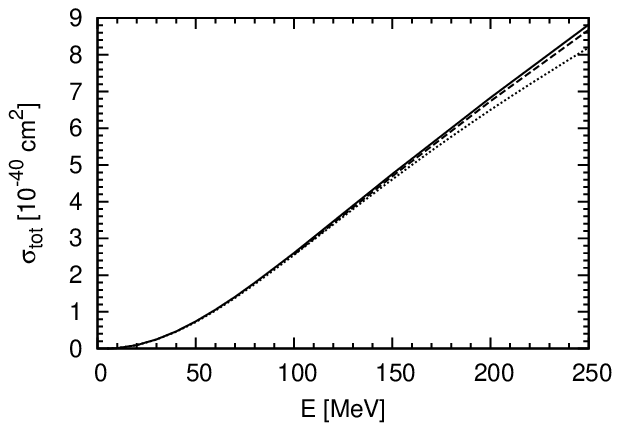}
\includegraphics[width=6cm]{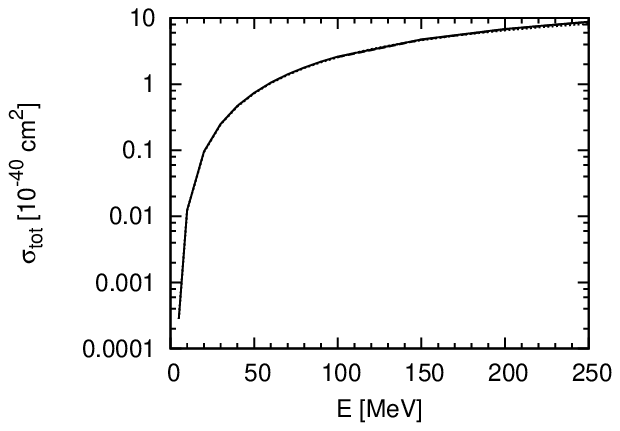}
\caption{The total cross section for the 
$\bar{\nu}_e + {^2{\rm H}} \rightarrow e^+ + n + n$
reaction  
as a function of the initial antineutrino energy $E$
shown on a linear (left panel) and on a logarithmic (right panel) scale.
The dashed (solid) line shows coordinate space predictions 
from Tabs.~II and IV of Ref.~\cite{PRC86.035503}
obtained with the AV18 potential and with the single nucleon current 
(with the inclusion of single- and 2N 
terms in the weak current operator), employing the relativistic 
kinematics. 
The dotted line displays 
results of nonrelativistic momentum space calculations
from the present work obtained with the AV18 potential and
with the single nucleon current.
\label{FIG.CCAANUE}}
\end{figure}

Let us start with the 
total cross section for the 
$\bar{\nu}_e + {^2{\rm H}} \rightarrow e^+ + n + n$
reaction, as it is the closest to the muon capture process 
studied in Ref.~\cite{PRC90.024001}. 
It is shown in Fig.~\ref{FIG.CCAANUE} as a function of the initial 
antineutrino energy $E$, both on a linear and on a logarithmic scale.
The dashed and solid lines in this figure 
show the predictions from Tabs.~II and IV of Ref.~\cite{PRC86.035503}
obtained with the AV18 potential and with the single-nucleon current
as well as with the single-nucleon current supplemented by the 2N 
currents linked to the AV18 potential.
The difference between these two curves highlights the 
importance of the 2N currents for this reaction. 
The third, dotted, line is used to display results of our nonrelativistic
calculations, carried out in momentum space with the AV18 potential and
with the single nucleon current.
Clearly for the energies $E \le$ 100~MeV all the three predictions essentially
overlap, but for higher energies effects of 2N contributions
to the current operator are visible. The relativistic 
treatment of the kinematics by Shen {\it et al.} 
leads to a clear spread between the dashed and the dotted line.

% FIG. 8

\begin{figure}
\includegraphics[width=6cm]{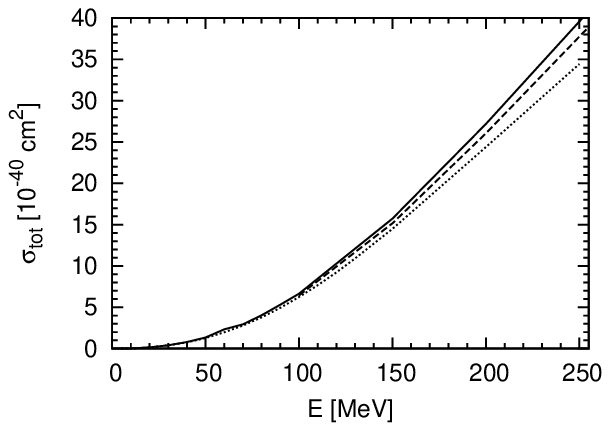}
\includegraphics[width=6cm]{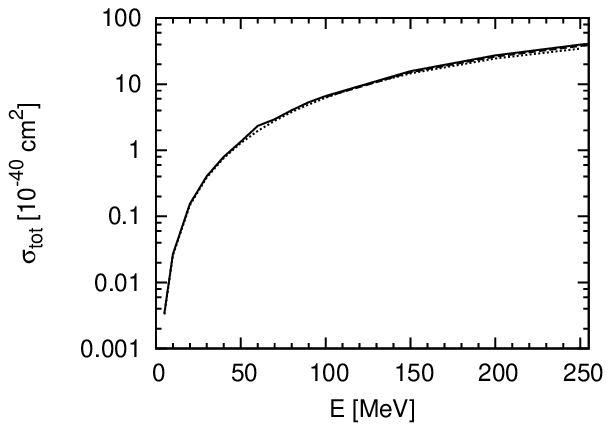}
\caption{The same as in Fig.~\ref{FIG.CCAANUE} for the 
$\nu_e + {^2{\rm H}} \rightarrow e^- + p + p$ reaction.
\label{FIG.CCNUE}}
\end{figure}

Next, in Fig.~\ref{FIG.CCNUE} we display in the same way results 
for the other CC driven reaction:
$\nu_e + {^2{\rm H}} \rightarrow e^- + p + p$.
Again our nonrelativistic predictions from momentum-space 
calculations are compared 
with the results from Ref.~\cite{PRC86.035503}. 
The values of the cross section are several times bigger 
and they rise with the neutrino energy at a different pace
than for the antineutrino induced reaction. While the scale of the 
relativistic effects in the kinematics is the same, the 2N 
currents seem to play a more important role for the highest energies considered 
here.

% FIG. 9

\begin{figure}
\includegraphics[width=6cm]{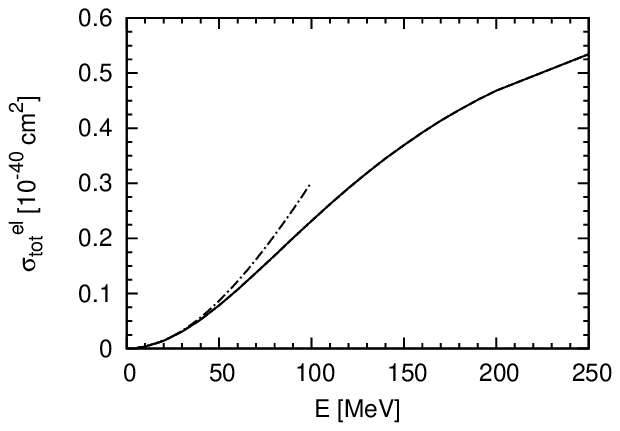}
\includegraphics[width=6cm]{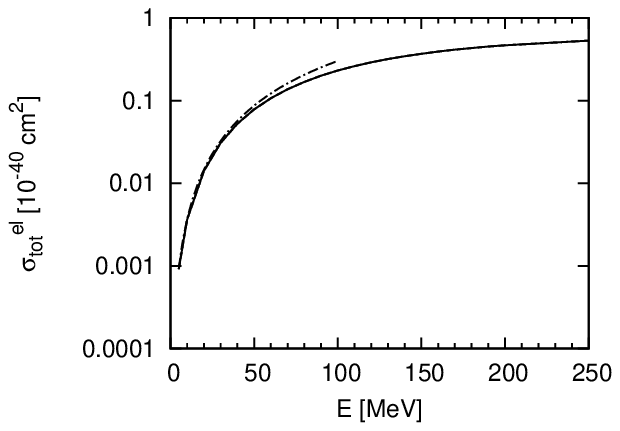}
\caption{The total cross section for the elastic NC (anti)neutrino scattering
off the deuteron as a function of the initial (anti)neutrino energy $E$,
shown on a linear (left panel) and on a logarithmic (right panel) scale.
The solid line displays 
results of nonrelativistic momentum space calculations
from the present work obtained with the AV18 potential and
with the single nucleon current.
These results are flavor independent and are the same for neutrino 
and antineutrino
scattering. In addition the dash-dotted line corresponds to the results from 
Butler {\it et al.}~\cite{Butler2000}, ignoring the strangeness contents 
of the deuteron. See text for more details.
\label{FIG.ELNC}}
\end{figure}

Elastic NC driven (anti)neutrino scattering on $^2$H
has not been considered in Refs.~\cite{PRC86.035503,Baroni17}.
It was studied, for example by 
Frederico {\it et al.} \cite{Frederico92}
and later by Butler {\it et al.} \cite{Butler2000},
who investigated also the neutrino-deuteron break-up reactions 
within a $\chi$EFT approach at next-to-leading order (NLO).
The authors of Ref.~\cite{Butler2000} derived analytical 
expressions for the elastic (anti)neutrino-deuteron scattering 
cross section, but did not yield direct results for the total cross sections.
They were interested in the effects caused by the presence of the strange
quarks in the deuteron. If the strangeness in the deuteron is 
neglected, the results for the elastic channel 
are not only flavor independent but just 
the same for neutrino and antineutrino scattering. 
That is exactly the case for our calculations presented
in Fig.~\ref{FIG.ELNC}. At the considered here low (anti)neutrino 
energies, this reaction is extremely hard to measure, due to the very small 
deuteron recoil energy. In addition, this reaction channel is strongly 
suppressed, as can be seen in Fig.~\ref{FIG.ELNC}, resulting 
in very small values of the total cross sections.
In Fig.~\ref{FIG.ELNC}, beside our predictions, we show also 
the results derived from Eq.~(34) in Ref.~\cite{Butler2000},
setting the strange form factors to zero and calculating 
the $F_C$ form factor using the simple lowest-order (LO) expression 
given by Eq.~(31) in that reference. For low (anti)neutrino energies, 
where the LO formula is valid, both types of results nicely agree
and clearly diverge in the higher-energy region.

% FIG. 10

\begin{figure}
\includegraphics[width=6cm]{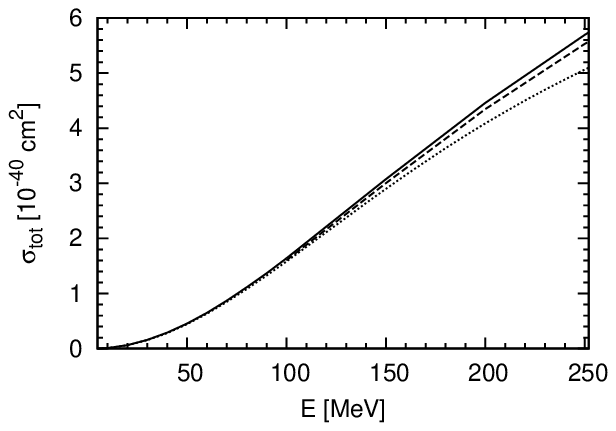}
\includegraphics[width=6cm]{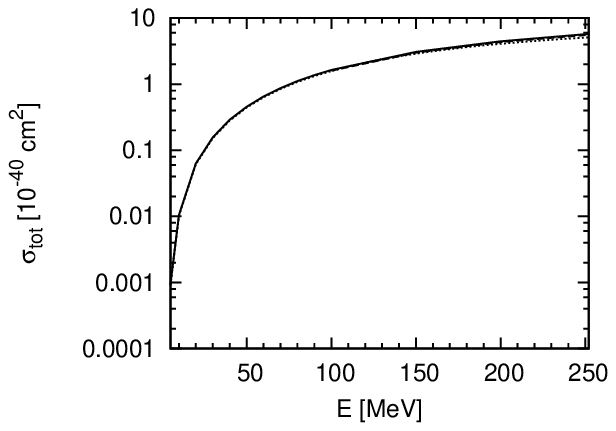}
\caption{The total cross section for the 
$\bar{\nu}_e + {^2{\rm H}} \rightarrow \bar{\nu}_e + p + n$
reaction  
as a function of the initial antineutrino energy $E$
shown on a linear (left panel) and on a logarithmic (right panel) scale.
The dashed (solid) line shows coordinate space predictions 
from Tabs.~II and III of Ref.~\cite{PRC86.035503}
obtained with the AV18 potential and with the single nucleon current 
(with the inclusion of single- and 2N 
terms in the weak current operator), employing the relativistic 
kinematics. 
The dotted line displays 
results of nonrelativistic momentum space calculations
from the present work obtained with the AV18 potential and
with the single nucleon current.
\label{FIG.NCAANUE}}
\end{figure}

\begin{figure}
\includegraphics[width=6cm]{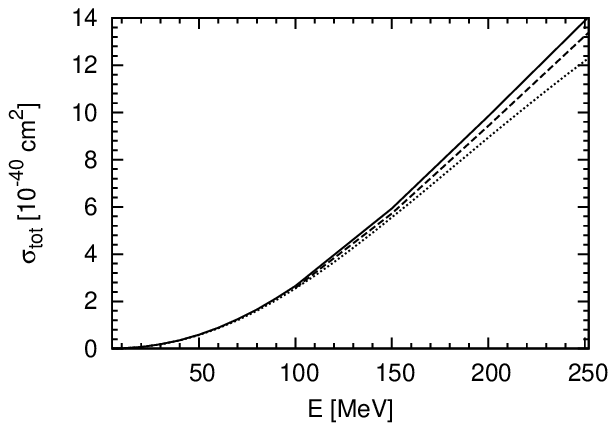}
\includegraphics[width=6cm]{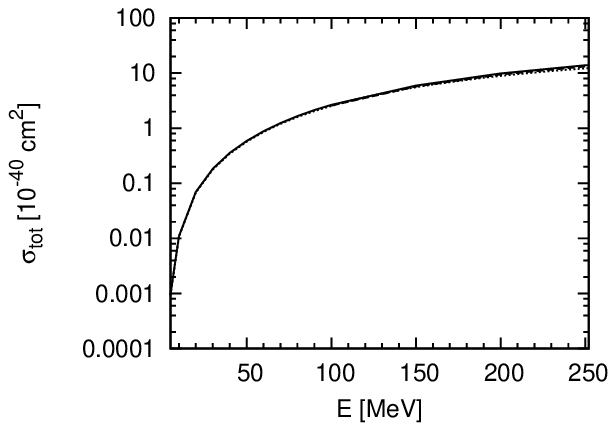}
\caption{The same as in Fig.~\ref{FIG.CCAANUE} for the 
$\nu_e + {^2{\rm H}} \rightarrow \nu_e + p + n$ reaction.
\label{FIG.NCNUE}}
\end{figure}

The inelastic NC induced reactions with the deuteron have been considered 
in Ref.~\cite{PRC86.035503} and in Figs.~\ref{FIG.NCAANUE} and \ref{FIG.NCNUE}
we again compare results based on coordinate-space and momentum-space
approaches, for the antineutrino and neutrino scattering, respectively.
As in Fig.~\ref{FIG.CCAANUE} and ~\ref{FIG.CCNUE},
we use results of Ref.~\cite{PRC86.035503}, now from Tabs.~II and~III, 
for the total cross section for the 
$\bar{\nu}_e + {^2{\rm H}} \rightarrow \bar{\nu}_e + p + n$
and
$\nu_e + {^2{\rm H}} \rightarrow \nu_e + p + n$ 
reactions, respectively.
The solid and dashed curves represent results 
from Ref.~\cite{PRC86.035503} with and without 2N 
contributions in the weak neutral nuclear current operator, accordingly.
They have been calculated with the relativistic kinematics,
so for higher energies in both figures the dashed lines visibly deviate 
from the dotted ones, representing our fully nonrelativistic 
momentum space predictions. These purely kinematical effects are of course essentially 
identical for the CC and NC driven reactions with electron (anti)neutrinos,
as the electron mass is very small compared to higher beam energies, 
where to a good approximation the electron could be treated 
as massless. The cross sections for the NC neutrino induced reaction 
are bigger than for the corresponding reaction with the antineutrinos
(roughly by a factor two), but this relative difference with respect 
to the antineutrino reaction is clearly smaller than for  
the CC induced reactions.

% FIG. 11

% FIG. 12

\section{Results for neutrino scattering on $^3$He and $^3$H}
\label{section4}

The results presented in this section have also been obtained with the single nucleon weak current operator 
from Refs.~\cite{Marcucci11,PRC86.035503,PRC90.024001} and with a nuclear Hamiltonian,
which contains only the 2N potential energy - the 3N force has been neglected. 

The same formalism which has been successfully developed for electron scattering and photodisintegration 
processes with 3N systems \cite{incRAB,raport2005} as well as for muon capture on the 3N bound states 
\cite{PRC90.024001,PRC94.034002} is directly applicable
to elastic, quasi-elastic and inelastic (anti)neutrino scattering on $^3$He and $^3$H.
The key elements of this formalism are presented in Sec.~\ref{section2}.

\begin{figure}
\includegraphics[width=0.4\textwidth,clip=true]{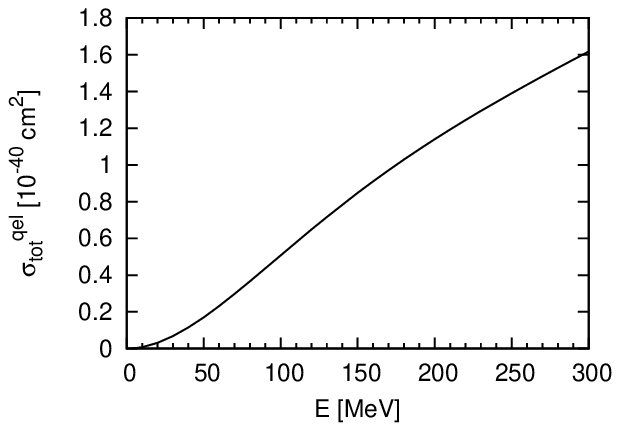}
\includegraphics[width=0.4\textwidth,clip=true]{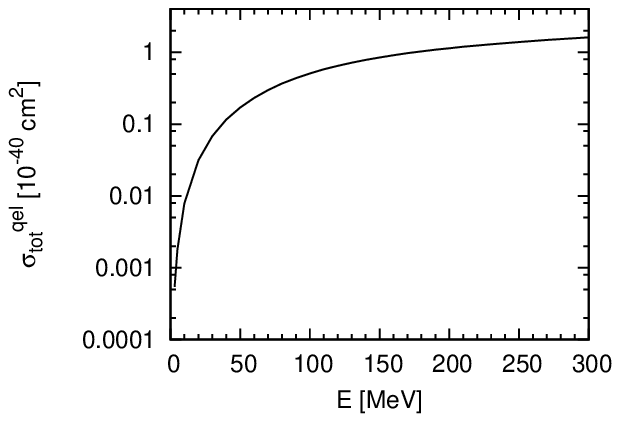}
\caption{The total cross section for the quasielastic CC 
$ \bar{\nu}_e + {^3{\rm He}} \rightarrow e^+ + {^3{\rm H}} $ process
as a function of the initial antineutrino energy $E$ 
shown on the linear (left panel) and logarithmic (right panel) scale.
The results are obtained with the AV18 NN potential and with the single nucleon CC operator,
which contains the relativistic corrections.
}
\label{fig4}
\end{figure}

First we consider the quasi-elastic electron antineutrino scattering on $^3$He, leading to
the positron and $^3$H nucleus in the final state. 
In this process the same as in muon capture weak CC operator
changes the total charge of the nuclear system. 
In Fig.~\ref{fig4} we show the total cross sections for the 
$ \bar{\nu}_e + {^3{\rm He}} \rightarrow e^+ + {^3{\rm H}} $
reaction as a function of the initial antineutrino energy.

% FIG. 13

\begin{figure}
\includegraphics[width=0.4\textwidth,clip=true]{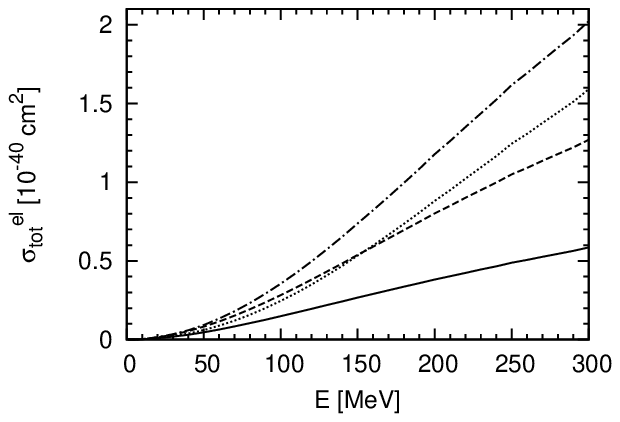}
\includegraphics[width=0.4\textwidth,clip=true]{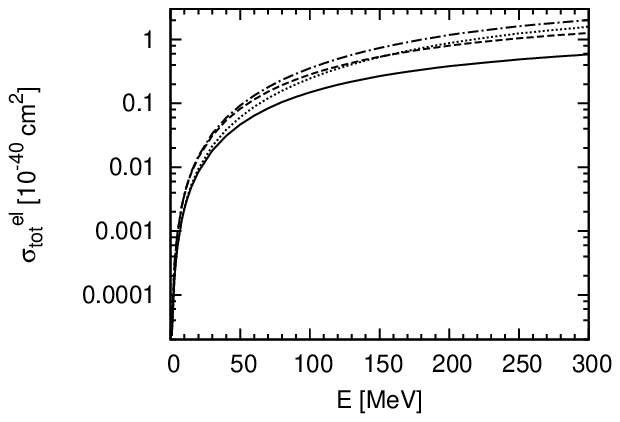}
\caption{The total cross section for the elastic NC driven reactions
$ \bar{\nu}_l + {^3{\rm He}} \rightarrow  \bar{\nu}_l + {^3{\rm He}} $  (solid line),
$ \bar{\nu}_l + {^3{\rm H}} \rightarrow  \bar{\nu}_l + {^3{\rm H}} $  (dashed line),
$ \nu_l + {^3{\rm He}} \rightarrow  \nu_l + {^3{\rm He}} $  (dotted line)
and
$ \nu_l + {^3{\rm H}} \rightarrow  \nu_l + {^3{\rm H}} $  (dash-dotted line)
as a function of the initial (anti)neutrino energy $E$ shown on a linear 
 (left panel) and a logarithmic (right panel) scale.
The results are obtained with the AV18 NN potential and the single nucleon NC operator,
which contains the relativistic corrections.
}
\label{fig44}
\end{figure}

Further non-break-up reactions with trinucleons are possible only with the 
neutral current. In Fig.~\ref{fig44} we display the predictions
for the total cross section of the elastic
$ \bar{\nu}_l + {^3{\rm He}} \rightarrow  \bar{\nu}_l + {^3{\rm He}} $,
$ \bar{\nu}_l + {^3{\rm H}} \rightarrow  \bar{\nu}_l + {^3{\rm H}} $,
$ \nu_l + {^3{\rm He}} \rightarrow  \nu_l + {^3{\rm He}} $,
and
$ \nu_l + {^3{\rm H}} \rightarrow  \nu_l + {^3{\rm H}} $ 
reactions as a function of the initial (anti)neutrino energy.
Despite very similar reaction kinematics, the cross sections take 
quite different values. The antineutrino-$^3$He cross section
is the smallest and the neutrino-$^3$H cross section reaches the highest values.
Lines representing neutrino-$^3$He and  antineutrino-$^3$H cross
sections cross at $E \approx$ 150 MeV, with the former prevailing 
for the higher energies. All the four predictions are flavor independent.

It is important to note that the difference between the predictions for 
the neutrino and antineutrino induced reactions on each nucleus 
comes solely from the replacement of $V_{ij}$ functions from 
Eq.~(\ref{V-fun}) by the $\bar{V}_{ij}$ functions defned in Eq.~(\ref{BARV-fun}).
In turn, the difference between the predictions for the 
$ \nu_l + {^3{\rm He}} \rightarrow  \nu_l + {^3{\rm He}} $
and
$ \nu_l + {^3{\rm H}} \rightarrow  \nu_l + {^3{\rm H}} $
reactions is caused nearly exclusively by the proton ($j_{NC}^p$) and neutron ($j_{NC}^n$) 
contributions in the following isospin matrix element:
\begin{eqnarray}
{\cal I}(t,m_T) \equiv 
\Big\langle  \left( t \frac12 \right) \frac12 \, m_T \mid 
\frac12 \left( 1 + \tau_{z} (1) \right) j_{NC}^p + \frac12 \left( 1 - \tau_{z} (1) \right) j_{NC}^n
\mid 
 \left( t \frac12 \right) \frac12 \, m_T \, \Big\rangle \, ,
\label{j1in3Nisospin}
\end{eqnarray}
which yield in the $^3$He case ($m_T=\frac12$) 
${\cal I} \left(0,\frac12 \right) = j_{NC}^p$,
${\cal I} \left(1,\frac12 \right) = \frac23 j_{NC}^n + \frac13 j_{NC}^p$,
and in the $^3$H case ($m_T=-\frac12$)
${\cal I} \left(0,-\frac12 \right) = j_{NC}^n$,
${\cal I} \left(1,-\frac12 \right) = \frac23 j_{NC}^p + \frac13 j_{NC}^n$.
The differences introduced by the slightly different masses 
and wave functions of $^3$He and $^3$H are practically negligible. The same is also true 
for the two antineutrino induced elastic reactions.

% FIG. 14

The CC and NC driven break-up reactions are definitely 
more demanding than the formerly discussed ones
due to the complicated kinematics, which has to take
into account two- and three-body disintegration processes. 
Full inclusion of the final state interactions is even more challenging,
especially for the isospin raising reactions induced by the neutrinos,
for which two or three outgoing nucleons are charged. 

In the studies of electron scattering on the trinucleons one usually
assumes that the initial electron energy, the electron scattering angle
and the final electron energy are known. This information allows one
to study the 3N scattering states at a fixed 3N internal energy 
and at a fixed total 3N momentum. In the case of the (anti)neutrino 
induced processes one is interested predominantly in the total cross
sections, which necessitates a calculation of at least hundreds of 
the neutrino kinematics. Even for the (anti)neutrino reactions 
with the deuteron calculations are indeed time consuming.  

\begin{figure}
\includegraphics[width=0.4\textwidth,clip=true]{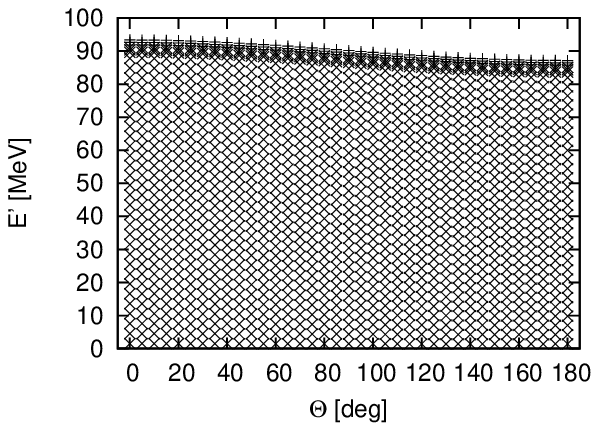}
\includegraphics[width=0.4\textwidth,clip=true]{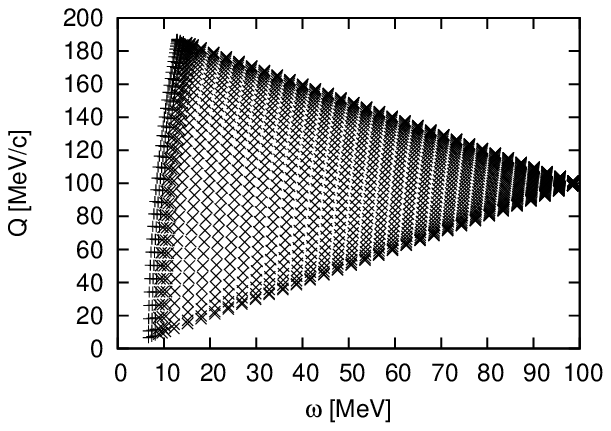}
\caption{The ranges of various kinematical quantities 
describing the kinematics of the NC driven break-up of $^3$H
for the initial (anti)neutrino energy $E$= 100 MeV. 
In the left panel the outgoing (anti)neutrino energy and the scattering
angle are considered, while in the right panel the magnitude of the three-momentum
transfer is plotted versus the energy transfer. 
}
\label{fig555}
\end{figure}

It is then important to realize that the essential dynamical
quantities for inclusive reactions, the so-called response functions $R_i$,
depend on two parameters only. These parameters are the energy transfer $\omega$
and the magnitude of the three-momentum transfer $Q \equiv \mid \bf Q \mid$. 
We show in Fig.~\ref{fig555} the ranges of these quantities for the NC neutrino 
inelastic scattering for the initial (anti)neutrino energy $E$= 100 MeV.
Of course the same statement is true also for the CC induced reactions,
so in both cases the total cross sections are built from the purely 
kinematical input and the response functions, calculated 
in the whole physical $\omega$-$Q$ region.

% FIG. 15

The response functions $R_{i} \equiv R_{i} (\omega, Q) $ 
stem from various products of the nuclear matrix elements: 
$R_{00} \propto \mid N^0 \mid^2 $,
$R_{ZZ} \propto \mid N_z \mid^2 $,
$R_{MM} \propto \mid N_{-1} \mid^2 $,
$R_{Z0} \propto {\rm Re} \left( N^0 ( N_z )^\star \, \right) $
and
$R_{PP} \propto {\mid N_{+1} \mid^2} $,
receiving contributions from all the final nuclear states.
Using the approach described in Eq.~(\ref{RAB4}), these contributions
can be separately calculated for the two 
values of the total 3N isospin, $T= \frac12$ and $T= \frac32$.
On the other hand the same sum over the final nuclear states
can be performed over the physical two-body and three-body 
fragmentation channels. An agreement between these two approaches 
provides a non-trivial test of numerics.

In this work we restrict ourselves to a sample of
results for the CC and NC response functions. 
They are calculated 
for the fixed value of the three-momentum transfer $Q$= 100 MeV/c,
as a function of the internal 3N energy $E_{c.m.}$. 
The latter quantity is simply related to the energy transfer.
For example, in the case of the antineutrino CC break-up of $^3$H
it reads 
\begin{eqnarray}
E_{c.m.} = \omega + M_{^3H} - 3 M_n - \frac{{\bf Q}^{\, 2}}{6 M_n} \, ,
\label{Ecm}
\end{eqnarray}
where the $M_{^3H}$ and $M_n$ are the triton and neutron masses,
respectively.

\begin{figure}
\includegraphics[width=0.4\textwidth,clip=true]{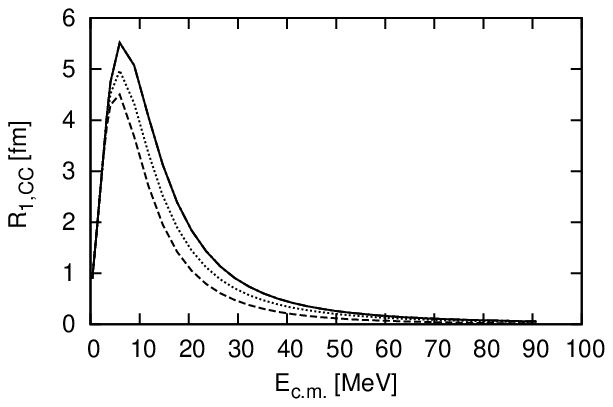}
\includegraphics[width=0.4\textwidth,clip=true]{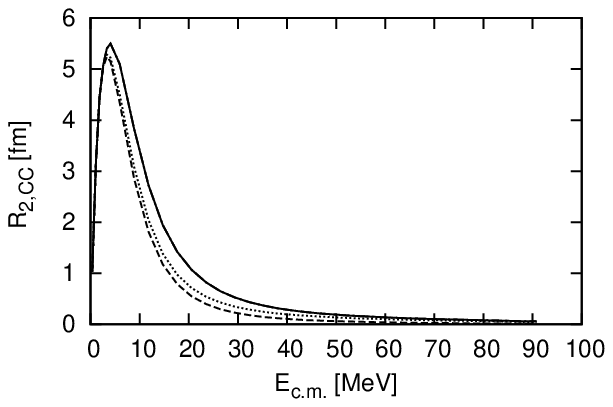}
\includegraphics[width=0.4\textwidth,clip=true]{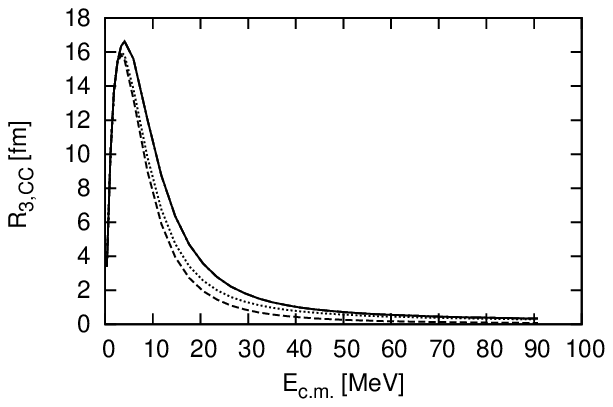}
\includegraphics[width=0.4\textwidth,clip=true]{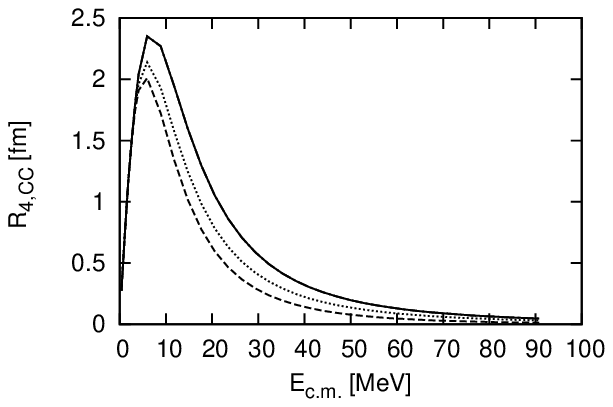}
\includegraphics[width=0.4\textwidth,clip=true]{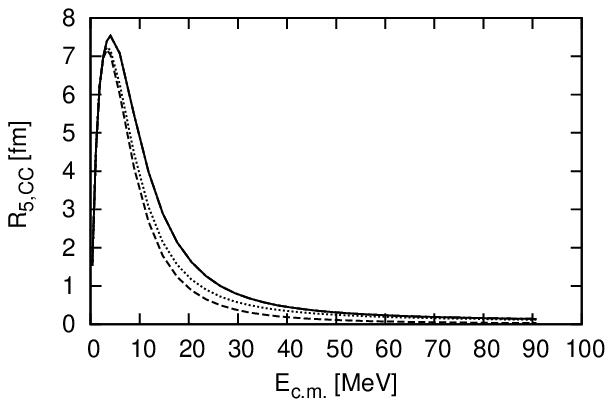}
\caption{The inclusive CC response functions  $R_{i,CC}$
for the electron antineutrino disintegration of $^3$He
as a function of the internal 3N energy $E_{c.m.}$ 
for the fixed value of the three-momentum transfer $Q$= 100 MeV/c.
The results are obtained with the AV18 NN potential 
and the single nucleon CC operator,
which contains the relativistic corrections.
The dotted line shows the contribution from final nuclear states
with the total isospin $T= \frac12$ and
the dash-dotted line represents the total response function
obtained from the total isospin $T= \frac12$ and $T= \frac32$ states.
The dashed line depicts the part of the response function stemming 
from the two-body break-up channel and the solid line the total response 
function, receiving contributions from both two- and three-body break-up
states. Note that the dash-dotted and solid lines overlap.
}
\label{fig444}
\end{figure}

The five inclusive CC response functions  $R_{i,CC}$
for the electron antineutrino disintegration of $^3$He
are shown in Fig.~\ref{fig444}. They all have a very similar shape,
known also from inclusive electron-nucleus scattering 
(see for example Ref.~\cite{Benhar07}):
they start from zero at threshold, rise to reach a maximum, whose 
position corresponds to antineutrino scattering elastically from a moving 
bound nucleon, and slowly tend to zero for higher $E_{c.m.}$ values.
In Fig.~\ref{fig444} we show
also separate contributions from the total isospin $T= \frac12$ 
states (dotted line) and from the two-body break-up channel (dashed line), 
while the total response function is computed either as a sum 
of the $T= \frac12$ and $T= \frac32$ parts (dash-dotted line) 
or as a sum over the two- and three-body break-up contributions (solid line).
As expected, for each $R_i$ the dash-dotted and solid lines overlap and the relative 
distance between the solid and dotted lines provides information
about the importance of the $T= \frac32$ 
(present only in the three-body break-up) contribution.
The corresponding difference 
between the solid and dashed lines provides information
about the contribution to the total response function 
from the three-body break-up channel.

% FIG. 16

\begin{figure}
\includegraphics[width=0.4\textwidth,clip=true]{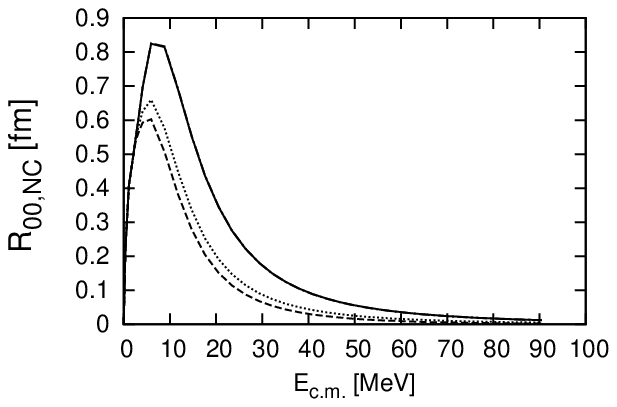}
\includegraphics[width=0.4\textwidth,clip=true]{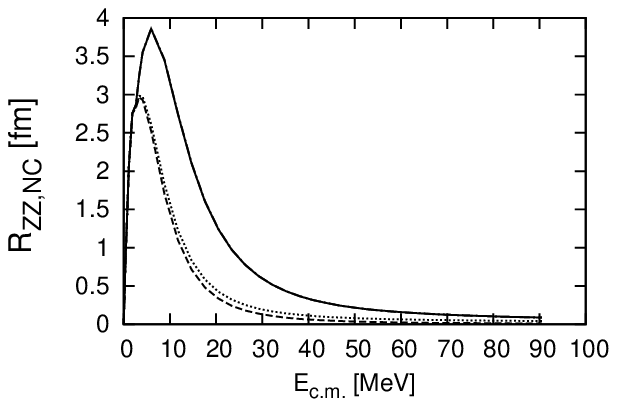}
\includegraphics[width=0.4\textwidth,clip=true]{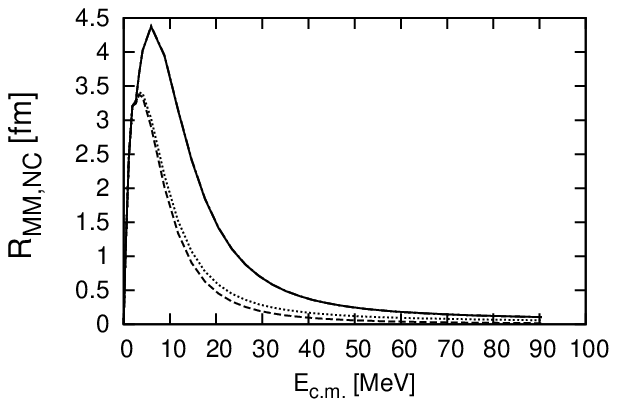}
\includegraphics[width=0.4\textwidth,clip=true]{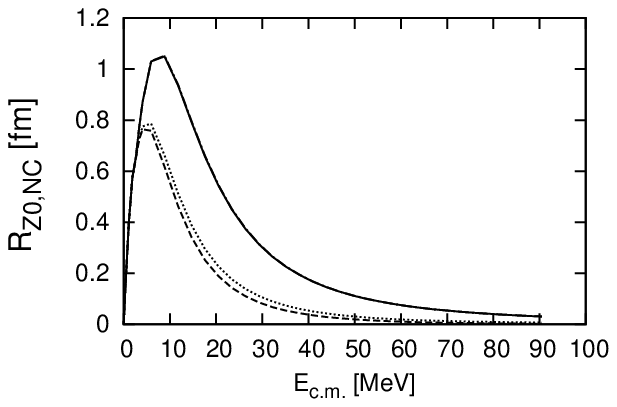}
\includegraphics[width=0.4\textwidth,clip=true]{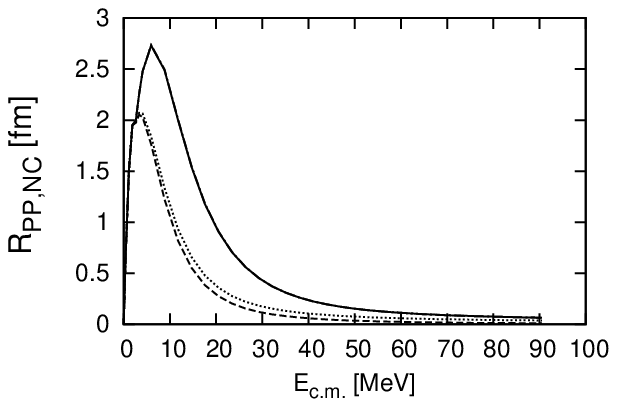}
\caption{The inclusive NC response functions $R_{i,NC}$
for the (anti)neutrino disintegration of $^3$H
as a function of the internal 3N energy $E_{c.m.}$
for the fixed value of the three-momentum transfer $Q$= 100 MeV/c.
The results are obtained with the AV18 NN potential
and the single nucleon NC operator,
which contains the relativistic corrections.
Lines as in Fig.~\ref{fig444}.
}
\label{fig666}
\end{figure}

\begin{figure}
\includegraphics[width=0.4\textwidth,clip=true]{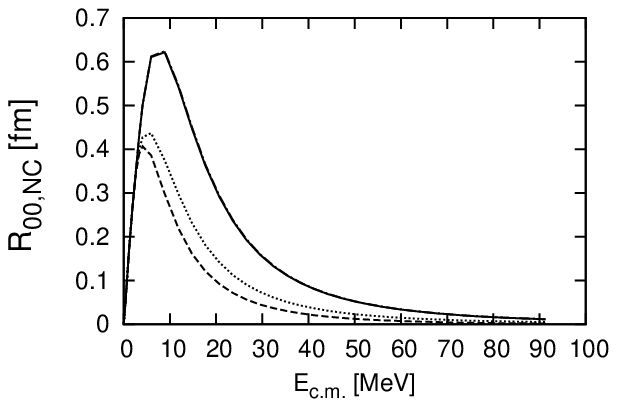}
\includegraphics[width=0.4\textwidth,clip=true]{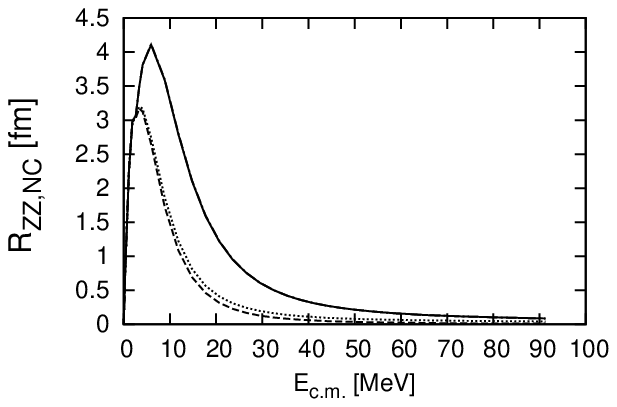}
\includegraphics[width=0.4\textwidth,clip=true]{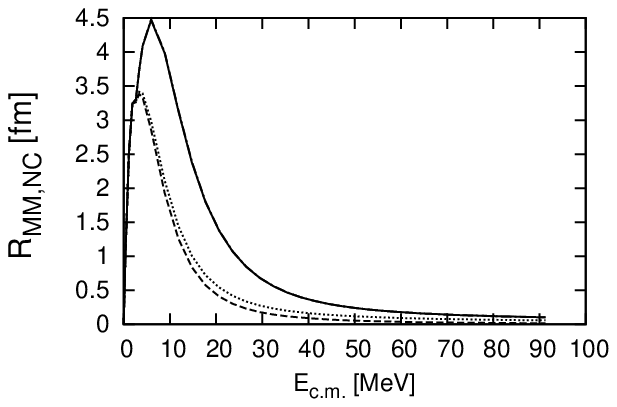}
\includegraphics[width=0.4\textwidth,clip=true]{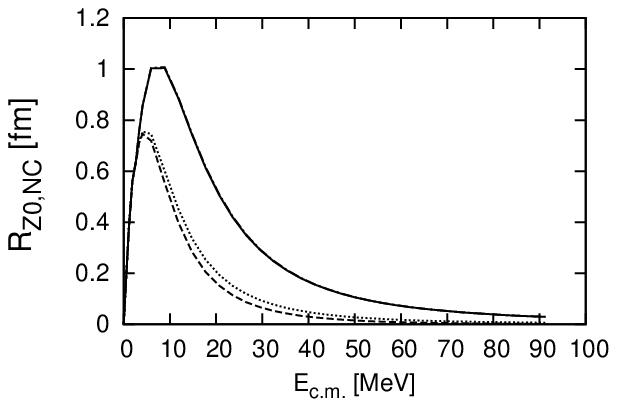}
\includegraphics[width=0.4\textwidth,clip=true]{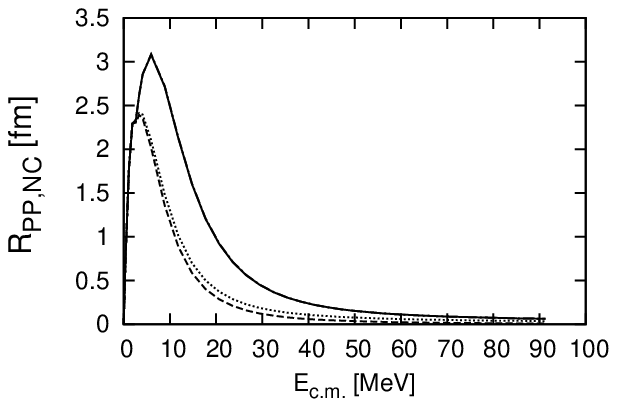}
\caption{The same as in Fig.~\ref{fig666} for the (anti)neutrino 
disintegration of $^3$He.
}
\label{fig777}
\end{figure}

In the same way we display the inclusive NC response functions $R_{i,NC}$
for the (anti)neutrino disintegration of $^3$H in Fig.~\ref{fig666}
and for the (anti)neutrino disintegration of $^3$He in Fig.~\ref{fig777}.
These response functions are the same for the neutrino and 
antineutrino induced reactions. From Figs.~\ref{fig666}
and \ref{fig777} we infer that both the $T= \frac32$ 
and the three-body break-up contributions to the inclusive NC response
functions are relatively much more important than for the
CC response functions. Our predictions 
for the (anti)neutrino disintegration of $^3$He suffer from the lack 
of Coulomb force in the calculations of the final nuclear states.
Our experience from investigations of electron scattering on $^3$He
tells us, however, that the Coulomb force does not
substantially affect {\em inclusive} response functions. 
Within our present framework we cannot describe CC neutrino 
induced processes with three protons in the final state.

% FIG. 17

% FIG. 18

\section{Summary and conclusions}
\label{section7}

A consistent framework 
for the calculations of several neutrino induced processes on
$^2$H, $^3$He, $^3$H and other light nuclei 
is still a challenge, despite the recent progress in this field. 
There are many models of the nuclear interactions and 
weak current operators linked to these forces, but
full compatibility has not been achieved yet.
The work on the regularization of the chiral potentials
and electroweak current operators is in progress
and the crucial issue of the current conservation has
to be ultimately solved. 
Additional problems arise from the fact that (anti)neutrinos 
can transfer a lot of energy and three-momentum to the nuclear 
system which requires a relativistic treatment of both kinematics and dynamics.

In the present paper we studied the two-nucleon 
$\bar{\nu}_e + {^2{\rm H}} \rightarrow e^+ + n + n$, 
$\nu_e + {^2{\rm H}} \rightarrow e^- + p + p$, 
$\bar{\nu}_l + {^2{\rm H}} \rightarrow \bar{\nu}_l + {^2{\rm H}}$,
$\nu_l + {^2{\rm H}} \rightarrow \nu_l + {^2{\rm H}}$,
$\bar{\nu}_l + {^2{\rm H}} \rightarrow \bar{\nu}_l + p + n$, 
$\nu_l + {^2{\rm H}} \rightarrow \nu_l + p + n$
and three-nucleon 
$\bar{\nu}_e + {^3{\rm He}} \rightarrow e^+ + {^3{\rm H}}$, 
$\bar{\nu}_l + {^3{\rm He}} \rightarrow \bar{\nu}_l + {^3{\rm He}}$, 
$\nu_l + {^3{\rm He}} \rightarrow \nu_l + {^3{\rm He}}$,
$\bar{\nu}_l + {^3{\rm H}} \rightarrow \bar{\nu}_l + {^3{\rm H}}$, 
$\nu_l + {^3{\rm H}} \rightarrow \nu_l + {^3{\rm H}}$,
$\bar{\nu}_e + {^3{\rm He}} \rightarrow e^+ + n + d$, 
$\bar{\nu}_e + {^3{\rm He}} \rightarrow e^+ + n + n + p$, 
$\bar{\nu}_l + {^3{\rm He}} \rightarrow \bar{\nu}_l + p + d$, 
$\bar{\nu}_l + {^3{\rm He}} \rightarrow \bar{\nu}_l + p + p +n$, 
$\nu_l + {^3{\rm H}} \rightarrow \nu_l + n + d$
and
$\nu_l + {^3{\rm H}} \rightarrow \nu_l + n + n + p$
reactions in the framework close to the one of Ref.~\cite{PRC86.035503} 
but with the single nucleon current operator.
For most of the reactions we provided predictions for the total cross sections. 
In the case of the (anti)neutrino-$^3$He 
and (anti)neutrino-$^3$H inelastic scattering we computed examples of the essential 
response functions.

The bulk of our results was obtained for the reactions with the deuteron.
Here, contrary to Ref.~\cite{PRC86.035503}, we restricted ourselves 
to the lower (anti)neutrino energies, where the use of our purely nonrelativistic
approach is better justified. But even in this more restricted range 
of neutrino energies relativistic effects in the kinematics were thoroughly
checked with the result that the main difference between the relativistic 
and nonrelativistic kinematics stemmed from the form 
of the phase space factor. 
We worked exclusively in momentum space, treating also the Coulomb 
interaction between the two outgoing protons in the form of a sharply cut off 
potential. This is justified because current matrix elements become
negligible for sufficiently large distances between two nucleons.
Last but not least, very important elements of the formalism - 2N scattering 
states in the partial wave representation - were cross-checked using the results 
from the momentum and coordinate space calculations.

The full understanding of the studied (anti)neutrino induced processes requires
the inclusion of at least 2N
contributions to the nuclear current operators and, especially at larger (anti)neutrino
energies, consistent relativistic treatment of kinematics and dynamics.
Note that momentum space offers an easier possibility to 
use the so-called ``boosted potential'', which has been already employed 
in various relativistic studies of few-nucleon systems~\cite{Kamada02,Witala05,Golak06,Liu08,Witala11,Polyzou11,Polyzou14}.
We plan to work along this line and perform more complete
calculations in the near future, ultimately with chiral dynamical input. 
We believe, however, that the predictions presented here 
can serve as an important benchmark.

\acknowledgments
This work is a part of the LENPIC project and was supported 
by the Polish National Science Center under Grants
No. 2016/22/M/ST2/00173 and 2016/21/D/ST2/01120. The numerical calculations were
partially performed on the supercomputer cluster of the JSC , J\"ulich, Germany.

\end{document}